\documentclass[%
 notitlepage,
 twocolumn,
 amsmath,amssymb,
 aps,
 pra,
]{revtex4-1}

\usepackage{graphicx}% Include figure files
\usepackage{dcolumn}% Align table columns on decimal point
\usepackage{bm}% bold math
\usepackage{ dsfont }
\newcommand{\revision}[1]{{ #1}}

\usepackage{color}
\begin{document}

\title{Adiabaticity when raising a uniform 3D optical lattice in a bimodal Bose-Einstein condensate}
\author{Dariusz Kajtoch,$^{1,\, 2}$  Emilia Witkowska$^{1}$ and Alice Sinatra$^{2}$}

\affiliation{$^{1}$Institute of Physics, PAS, Aleja Lotnikow 32/46, PL-02668 Warsaw, Poland\\
$^{2}${Laboratoire Kastler Brossel, ENS--Universit\'e PSL, CNRS, 
   Sorbonne Universit\'e, Coll\`ege de France,
   24 rue Lhomond, 75005 Paris, France}}
\date{\today}

\begin{abstract}
Using the time-dependent Bogoliubov approach, we study adiabaticity for a two-component Bose-Einstein condensate in a 3D time-dependent optical lattice with unit filling,
in the superfluid and weakly interacting regime. We show that raising the lattice potential height can couple the ground state of the Bogoliubov Hamiltonian to excited states with two quasiparticles of opposite quasi-momenta. In the symmetric case {for interactions and density in the two components} these represent sound waves where the two components oscillate out of phase. We find an analytic expression of the adiabatic time, its dependence on the fraction of atoms in each component and its scaling with the system size.
\end{abstract}

\maketitle

% ===========================================
% Section 1
% ===========================================
\section{Introduction}
\label{sec:intro}

Spinor bosonic or fermionic atoms in optical lattices are playing an increasingly important role at the crossing of different fields {such as} statistical physics, condensed matter, atomic physics and quantum technologies.
{Actively investigated in present experiments, these systems display non trivial phase transitions and ground states stemming from the interplay between the spin and the external degrees of freedom of the atoms \cite{Demler}, and  can be used to investigate novel superfluidity mechanisms 
\cite{Hofstetter}. Furthermore, they constitute a powerful platform for quantum computation \cite{Briegel,PNAS}, and offer fascinating perspectives for entangled state preparation and quantum metrology
\cite{Jaksch,AtomicCrystal}. 
Among the different proposals using cold atoms in an optical lattice,  
several protocols require the possibility to {\it adiabatically} ramp up the optical lattice in the multi-component cold atoms system.} 

{In close relation with the first experimental realizations}, adiabaticity has been mainly studied, both experimentally \cite{Gericke} and theoretically \cite{KubaPRA,KubaDelande} for a single species and in the presence of an external harmonic potential that brakes the translational symmetry of the lattice.
In this paper we concentrate on the case of a uniform optical lattice, that is now possible to prepare in the laboratory thanks to the development of flat-bottom potentials \cite{Hadzibabic,Zwirlein}. 
We shall consider a single-component or a two-component condensate in the superfluid regime and study the adiabaticity condition when raising the lattice within the time-dependent Bogoliubov approach.

For a quantum system with a discrete spectrum $\hat{H}\Psi_k=E_k\Psi_k$, initially in its ground state $\Psi_0$, the adiabaticity condition when a parameter of the Hamiltonian is varied in time starting from $t=0$
takes the form \cite{Messiah2003}
\begin{equation}
\label{eq:adiabatic_cond}
\hbar  \frac{ \left| \langle\Psi_k(t)|\frac{d}{dt} \Psi_0(t)\rangle \right|}{|E_{k}(t)-E_0(t)|}   \ll 1 \quad \forall  k \neq 0,
\quad \forall  t>0
\end{equation}
$\langle\Psi_k(t)|\frac{d}{dt} \Psi_0(t)\rangle$ is the coupling between the instantaneous ground state and other eigenstates, 
and $E_k(t) - E_0(t)$ is the energy difference. The condition (\ref{eq:adiabatic_cond}) can be equivalently rewritten as
\begin{equation}
\label{eq:adiabatic_cond2}
 \frac{ \left| \langle\Psi_k(t)  |  i\hbar \frac{d\hat{H}}{dt}| \Psi_0(t)\rangle \right|}{|E_{k}(0)-E_0(t)|^2}  \ll 1 \quad \forall  k \neq 0,
\quad \forall  t>0 
\end{equation}
We will use this last formulation to interpret our result. 
Our paper is structured as follows: after recalling the Bose-Hubbard model and introducing all notations, the adiabaticity criterion for a single component Bose-Einstein condensate is derived in section \ref{sec:onecomp}.
The analysis is extended to the case of a two-components system in section \ref{sec:twocomp}. \revision{While in this section we keep our formalism general, allowing to address the case where the atoms in the two components might have different masses, in the following we concentrate on the case of equal masses corresponding to different hyperfine states of the same atomic species. For this case, in section \ref{sec:equalmasses}, we derive the adiabatic time for a linear ramp and we 
investigate the influence of a density imbalance between the two components.}
Conclusions are drown in section \ref{sec:concl}.

%-------------------------------------------------------------------------------------------
\section{Adiabaticity criterion for a single component}
\label{sec:onecomp}
%-------------------------------------------------------------------------------------------
In the adiabatic evolution, when an external parameter of the Hamiltonian is changed in time, the quantum state remains an instantaneous eigenstate of the time-dependent Hamiltonian at all times. On the contrary, if the change is too fast, the state will contain an admixture of excited states. For a Bose-Einstein condensate in an optical lattice in the weakly interacting regime, when the system is superfluid, the excitations are well described as Bogoliubov quasi-particles. We will regard the evolution as adiabatic as long as the total density of excited quasi-particles remains much less than one at all times. 

\subsection{Single component Bose-Hubbard model}
We describe a system of $N$ ultra-cold atoms, all in the same internal state and subject to an optical lattice potential, by the Bose-Hubbard Hamiltonian \cite{PhysRevLett.81.3108}
\begin{equation}\label{eq:ham}
\hat{\mathcal{H}}(t) = -J(t) \sum\limits_{\langle i,j \rangle} \hat{a}_i^{\dagger}(t) \hat{a}_j(t) + \frac{U(t)}{2} \sum\limits_i \hat{n}_i(t) [ \hat{n}_i(t) - 1],
\end{equation}
where $\hat{a}_i^{\dagger}(t)$ creates a particle in the single-particle Wannier state $w_i(\mathbf{r},t)$ of the lowest energy band ($l=1$) localized on the $i$-th site. The Bose-Hubbard model~(\ref{eq:ham}) considers only states in the lowest energy band, which is justified as long as the excitations energies to the higher bands are much larger {than} the energies involved in the system dynamics. 
The Wannier states $w_i(\mathbf{r},t)$ are conveniently constructed from the lowest band Bloch states $\psi_{l=1,\mathbf{q}}(\mathbf{r},t)$ in the following way
\begin{equation}\label{eq:wannier_bloch}
w_i(\mathbf{r},t) = \frac{1}{\sqrt{M}} \sum\limits_{\mathbf{q} \in BZ} e^{-i \mathbf{q} \mathbf{R}_\mathbf{i}} \psi_{l=1,\mathbf{q}}(\mathbf{r},t),
\end{equation}
with $\mathbf{R}_\mathbf{i} = d(i_x \mathbf{u}_x + i_y \mathbf{u}_y + i_z \mathbf{u}_z)$, where $d=\lambda/2$ is the lattice spacing that we assume identical in the three spatial directions ($\lambda$ is the optical lattice wavelength), $\mathbf{u}_{\alpha}$ are unit vectors and $i_x,i_y,i_z$ are integers. Summation in Eq.~\eqref{eq:wannier_bloch} extends over wave vectors belonging to the 1st Brillouin zone and $M$ is the number of lattice sites. The Bloch states $\psi_{l,\mathbf{q}}(\mathbf{r},t)$, labeled by the band index $l$ and the quasi-momentum $\mathbf{q}$, are eigenstates of the single-particle Hamiltonian
\begin{equation}\label{eq:single_particle_hamiltonian}
\hat{h}(t) = -\frac{\hbar^2}{2 m} \nabla^{2} + V_0(t) \sum\limits_{\alpha = x,y,z} \sin^{2}(k \alpha), 
\end{equation}
where $k = 2\pi/\lambda$ and $V_{0}(t)$ is the lattice potential height. 
If $V_0(t)$ is varied in time,
the Wannier states and hence
the creation and annihilation operators in the Bose-Hubbard Hamiltonian~\eqref{eq:ham}, as well as the hoping $J(t)$ and interaction $U(t)$ parameters, 
\begin{subequations}\label{eq:hoping_inter} depend on time
 \begin{align}
  U(t) & = \frac{4 \pi a_s \hbar^2}{m} \int d^3 r\ |w(\mathbf{r},t)|^{4} , \label{eq:u_definition}\\\
  J(t) & = \int d^{3}r\ w^{*}_{i}(\mathbf{r},t) \left[-\frac{\hbar^2}{2 m}\nabla^2\right. \label{eq:J} \nonumber\\
  &\left. + V_0(t) \sum\limits_{\alpha = x,y,z} \sin^{2}(k \alpha) \right] w_j(\mathbf{r},t),
 \end{align}
\end{subequations}
where $a_s$ is the $s$-wave scattering length characterizing binary short range interactions between cold atoms and $m$ is the mass of an atom. In the limit in which $V_0\gg E_R$ where 
$E_R=\hbar^2k^2/(2m)$ is the recoil energy, the dependence of $U$ and $J$ on the lattice height $V_0$ can be approximated by~\cite{RevModPhys.80.885,Zwerger2003}
\begin{subequations}
 \begin{align}
  U(V_0) & = E_R \sqrt{\frac{8}{\pi}}k a_s \left(\frac{V_0}{E_R}\right)^{3/4}, \label{eq:GaussU}\\
  J(V_0) & = E_R \frac{4}{\sqrt{\pi}} \left(\frac{V_0}{E_R}\right)^{3/4} e^{-2\sqrt{\frac{V_0}{E_R}}}. \label{eq:GaussJ}
 \end{align}
\end{subequations}
For brevity, we shall omit in the following to mark the band index $l=1$ for the Bloch states of the lowest band.

%-------------------------------------------------------------------------------------------
\subsection{Number-conserving Bogoliubov approach}
We start from the Bose-Hubbard Hamiltonian~\eqref{eq:ham} written in the quasi-momentum representation
\begin{equation}\label{eq:single_ham_momentum}
\hat{\mathcal{H}}(t) = \sum\limits_{\mathbf{q}} \epsilon_{\mathbf{q}}(t) \hat{a}^{\dagger}_{\mathbf{q}}\hat{a}_{\mathbf{q}} + \frac{U(t)}{2M}\sum\limits_{\mathbf{q}_1, \mathbf{q}_2, \mathbf{k}}\hat{a}^{\dagger}_{\mathbf{q}_1 - \mathbf{k}}\hat{a}^{\dagger}_{\mathbf{q}_2 + \mathbf{k}}\hat{a}_{\mathbf{q}_1}\hat{a}_{\mathbf{q}_2},
\end{equation}
contrarily to the homogeneous case, the kinetic energy in the lattice takes the form 
\begin{equation}
 \epsilon_{\mathbf{q}}(t) = -2J(t)\sum_{\alpha=x,y,z}\cos(\mathbf{q} \cdot d\mathbf{u}_\alpha).
 \label{eq:epsilonq}
\end{equation}
In Eq.~\eqref{eq:single_ham_momentum} and further we omit the explicit time dependence of the creation and annihilation operators to simplify the notation, but we keep in mind that, 
even in the Schr\"{o}dinger picture, $\hat{a}^{\dagger}_{\mathbf{q}}$ and $\hat{a}_{\mathbf{q}}$ depend on time.

In the number conserving Bogoliubov approach the amplitude of the field in the condensate mode is finally eliminated, and the interacting system is described as an ensemble of quasiparticles in the modes orthogonal to the condensate mode. The small parameter of the theory is the non-condensed fraction, and to the lowest non-zero order, that is the Bogoliubov order, the quasiparticles do not interact. 
The first step to find the Bogoliubov quasi-particles is to quadratize the Hamiltonian with respect to the non-condensed field. One then obtains
\begin{align}\label{eq:bogoliubov}
& \hat{\mathcal{H}}(t) \simeq N \left[ \epsilon_\mathbf{0}(t) + \frac{U(t)}{2} n \right] + \sum\limits_{\mathbf{q}\neq \mathbf{0}} \hat{a}^{\dagger}_{\mathbf{q}}\hat{a}_{\mathbf{q}} \left[\epsilon_\mathbf{q}(t) - \mu(t) \right] \nonumber \\
& + \frac{U(t)}{2M}\sum\limits_{\mathbf{q}\neq \mathbf{0}} \left[ \hat{a}^{\dagger}_{\mathbf{0}}\hat{a}^{\dagger}_{\mathbf{0}} \hat{a}_{\mathbf{q}}\hat{a}_{-\mathbf{q}} + \hat{a}_{\mathbf{0}}\hat{a}_{\mathbf{0}} \hat{a}^{\dagger}_{\mathbf{q}}\hat{a}^{\dagger}_{-\mathbf{q}} 
+ 4 N  \hat{a}^{\dagger}_{\mathbf{q}}\hat{a}_{\mathbf{q}} \right]
\end{align}
where $n=N/M$ is the total atom density and the chemical potential is defined as
\begin{equation}
 \mu(t) = \epsilon_\mathbf{0}(t) + U(t)n.
\end{equation}
To obtain Eq.~\eqref{eq:bogoliubov}, the relation $\hat{a}_{\mathbf{0}}^{\dagger}\hat{a}_{\mathbf{0}} = N - \delta \hat{N}$, where   $\delta \hat{N} = \sum_{\mathbf{q} \neq \mathbf{0}}\hat{a}^{\dagger}_{\mathbf{q}}\hat{a}_{\mathbf{q}}$, was used.
We now introduce the number conserving operators 
\begin{equation}
\hat{\Lambda}_{\mathbf{q}} = \frac{1}{\sqrt{N}}\hat{a}^{\dagger}_{\mathbf{0}}\hat{a}_{\mathbf{q}} \label{eq:Lambda}
\end{equation}
in terms of which the Bogoliubov Hamiltonian~\eqref{eq:bogoliubov} takes its final form
\begin{equation}
\hat{\mathcal{H}}_{\rm Bog}(t) = H_0(t) + \frac{1}{2}\sum\limits_{\mathbf{q}\neq \mathbf{0}} 
\left( \hat{\Lambda}^{\dagger}_{\mathbf{q}}, \hat{\Lambda}_{-\mathbf{q}}\right) 
\sigma_z \mathcal{L}_{\mathbf{q}}
\left( \begin{array}{c}
\hat{\Lambda}_{\mathbf{q}} \\
\hat{\Lambda}^{\dagger}_{-\mathbf{q}}
\end{array}\right),
\label{eq:one_bogoliubov_matrix}
\end{equation}
where the ground state energy is
\begin{equation}
 H_0(t) = N \left[ \epsilon_\mathbf{0}(t) + \frac{U(t)}{2}n \right] - \frac{1}{2}\sum\limits_{\mathbf{q}\neq \mathbf{0}}\left[ \epsilon_{\mathbf{q}}(t) -\mu(t) + 2U(t)n \right],
 \label{eq:onecomp_H0}
\end{equation}
the non-hermitian matrix $\mathcal{L}_{\mathbf{q}}$ has the form
\begin{equation}
\mathcal{L}_{\mathbf{q}} = 
\left( 
 \begin{array}{cc}
  \epsilon_{\mathbf{q}}(t) -\mu(t) + 2U(t)n & U(t)n  \\
  -U(t)n &  -\left[ \epsilon_{\mathbf{q}}(t) -\mu(t) + 2U(t)n\right]
 \end{array}
\right)\,,
\label{eq:onecomp_L}
\end{equation}
and $\sigma_z$ is the third Pauli matrix.
In the derivation of (\ref{eq:one_bogoliubov_matrix}) we used {the following approximation}
\begin{equation}
\hat{\Lambda}^{\dagger}_{\mathbf{k}} \hat{\Lambda}_{\mathbf{q}} = \frac{\hat{a}^{\dagger}_{\mathbf{k}} \hat{a}_{\mathbf{q}} \hat{a}_{\mathbf{0}} \hat{a}^{\dagger}_{\mathbf{0}}}{N} \approx  \hat{a}^{\dagger}_{\mathbf{k}} \hat{a}_{\mathbf{q}},
\end{equation}
{consistent with the fact that} we retain only terms that are at most quadratic in the non-condensed field.
The Bogoliubov transformation $\mathcal{T}_{\mathbf{q}}(t)$
\begin{align}\label{eq:bogoliubov_single}
\left(
 \begin{array}{c}
  \hat{b}_{\mathbf{q}} \\
  \hat{b}^{\dagger}_{-\mathbf{q}}
 \end{array}
\right) 
=
\underbrace{%
\left( 
 \begin{array}{cc}
  \bar{u}_{\mathbf{q}}(t) & -\bar{v}_{\mathbf{q}}(t) \\
  -\bar{v}_{\mathbf{q}}(t) & \bar{u}_{\mathbf{q}}(t)
 \end{array}
\right)
}_{\mathcal{T}_{\mathbf{q}}(t)}
\left(
 \begin{array}{c}
  \hat{\Lambda}_{\mathbf{q}} \\
  \hat{\Lambda}^{\dagger}_{-\mathbf{q}}
 \end{array}
\right),
\end{align}
diagonalizes the quadratic Hamiltonian \eqref{eq:one_bogoliubov_matrix} in terms of the Bogoliubov operators $\hat{b}_{\mathbf{q}}$ {and $\hat{b}_{\mathbf{q}}^\dagger$} that satisfy bosonic commutation relations
$[\hat{b}_{\mathbf{q}},\hat{b}_{\mathbf{q'}}^\dagger]=\delta_{\mathbf{q},\mathbf{q'}}$
\begin{align}
\hat{\mathcal{H}}_{\rm Bog}(t) = & H_0 + \frac{1}{2}\sum\limits_{\mathbf{q}\neq \mathbf{0}}\hbar \omega_{\mathbf{q}}(t) + \sum\limits_{\mathbf{q}\neq \mathbf{0}} \hbar \omega_{\mathbf{q}}(t) \hat{b}^{\dagger}_{\mathbf{q}} \hat{b}_{\mathbf{q}},
\label{eq:HBog}
\end{align}
where the Bogoliubov energy has the form 
\begin{align}
\hbar \omega_{\mathbf{q}}(t) & = \sqrt{ \Delta E_{\mathbf{q}}(t)  \left[ \Delta E_{\mathbf{q}}(t) + 2 \tilde{\mu}(t) \right]} \,, \label{eq:bogoliubov_energy} \\
\Delta E_{\mathbf{q}}(t) &= \epsilon_{\mathbf{q}}(t) - \epsilon_{\mathbf{0}}(t) \,, \label{eq:DeltaE}\\
\tilde{\mu}(t) & = \mu(t) - \epsilon_{\mathbf{0}}(t) = U(t)n\,,
\end{align}
and the Bogoliubov modes are

{
\begin{align}
\bar{u}_{\mathbf{q}}(t)+\bar{v}_{\mathbf{q}}(t) &=  \left( \frac{\Delta E_{\mathbf{q}}(t)}{\Delta E_{\mathbf{q}}(t) + 2\tilde{\mu}(t)}\right)^{1/4} \,,\\
\bar{u}_{\mathbf{q}}(t)-\bar{v}_{\mathbf{q}}(t) &=  \left( \frac{\Delta E_{\mathbf{q}}(t)}{\Delta E_{\mathbf{q}}(t) + 2\tilde{\mu}(t)}\right)^{-1/4} \;. \label{eq:bogoliubov_modes} 
\end{align}
}

In the limit of {a small quasi-momentum $dq\ll1$ and for $\Delta E_{\mathbf{q}}\ll U(t)n$,} the Bogoliubov energy \eqref{eq:bogoliubov_energy} has a phonon-like dispersion
\begin{align}
\hbar \omega_{\mathbf{q}}(t) & = \hbar c(t) q
\end{align}
where $c(t) = a \sqrt{2 J(t) U(t) n}/\hbar$ is the sound velocity. The Bogoliubov spectrum for a condensate in a uniform lattice was already found in \cite{Burnett} within the usual symmetry-breaking approach.

 %-------------------------------------------------------------------------------------------
\subsection{Time evolution in the Heisenberg picture}

The column vector composed of Bogoliubov quasi-particle annihilation and creation operators
\begin{align}
\hat{B}_{\mathbf{q}}(t) = & \left( 
\begin{array}{c}
\hat{b}_{\mathbf{q}}(t) \\
\hat{b}_{-\mathbf{q}}^{\dagger}(t)
\end{array}
\right),
\end{align}
evolves according to the Heisenberg equation of motion
\begin{equation}\label{eq:HeisenbergforB}
\frac{d}{dt} \hat{B}_{\mathbf{q}}(t) = \frac{i}{\hbar}[\hat{\mathcal{H}}_{\rm Bog}(t), \hat{B}_{\mathbf{q}}(t)] + \left( \frac{\partial \hat{B}_{\mathbf{q}}(t)}{\partial t}\right)_{H}.
\end{equation}

The first part on the right-hand side of~\eqref{eq:HeisenbergforB} represents the free evolution of the quasi-particles
\begin{equation}
[\hat{\mathcal{H}}_{\rm Bog}(t), \hat{B}_{\mathbf{q}}(t)] = -\omega_{\mathbf{q}}(t) \sigma_z \hat{B}_{\mathbf{q}}(t) \,.
\end{equation}
 
According to the Bogoliubov transformation (\ref{eq:bogoliubov_single}),
the second part on the right-hand side of~\eqref{eq:HeisenbergforB} reads
\begin{equation}\label{eq:split}
\left( \frac{\partial \hat{B}_{\mathbf{q}}(t)}{\partial t}\right)_{H} = \frac{d \mathcal{T}_{\mathbf{q}}(t)}{d t} \mathcal{T}_{\mathbf{q}}^{-1}(t) \hat{B}_{\mathbf{q}}(t) + \mathcal{T}_{\mathbf{q}}(t) \partial_t \left( \begin{array}{c}
\hat{\Lambda}_{\mathbf{q}}(t) \\
\hat{\Lambda}_{-\mathbf{q}}^{\dagger}(t) 
\end{array}\right)
\end{equation}
By using the identity $\bar{u}_{\mathbf{q}}^2(t) - \bar{v}_{\mathbf{q}}^2(t) = 1$ one can show that 
\begin{equation}
\frac{d \mathcal{T}_{\mathbf{q}}(t)}{d t} \mathcal{T}_{\mathbf{q}}^{-1}(t) =  -\Omega_{\mathbf{q}}(t) \sigma_{x},
\end{equation}
where
\begin{equation}
\Omega_{\mathbf{q}}(t) =  \bar{u}_{\mathbf{q}}(t)\frac{d}{dt}\bar{v}_{\mathbf{q}}(t) - \bar{v}_{\mathbf{q}}(t)\frac{d}{dt}\bar{u}_{\mathbf{q}}(t), \label{eq:coupling}
\end{equation}
and $\sigma_x$ is the first Pauli matrix. 
The coupling $\Omega_{\mathbf{q}}(t)$ can be expressed in terms of the quasi-particles energies, and takes the form
\begin{equation}\label{eq:one_omega}
\Omega_{\mathbf{q}}(t) = \frac{d}{dt}\log(\bar{u}_{\mathbf{q}} + \bar{v}_{\mathbf{q}}) = \frac{1}{2}\frac{d}{dt}\log\left( \frac{\Delta E_{\mathbf{q}}(t)}{\hbar \omega_{\mathbf{q}}(t)} \right).
\end{equation}
Let us now deal with the second term on the right-hand side of Eq.~\eqref{eq:split}. 
The time derivative of the number-conserving operator (\ref{eq:Lambda}) is
\begin{equation}\label{eq:single_lambda_derivative}
\partial_t \hat{\Lambda}_{\mathbf{q}}(t) = \frac{1}{\sqrt{N}} \left[ \frac{\partial \hat{a}^{\dagger}_{\mathbf{0}}(t)}{\partial t} \hat{a}_{\mathbf{q}}(t) + \hat{a}^{\dagger}_{\mathbf{0}}(t) \frac{\partial \hat{a}_{\mathbf{q}}(t)}{\partial t} \right]\,.
\end{equation}
By using the definition of the creation and annihilation operators in the first Bloch band
\begin{equation}
  \hat{a}_{\mathbf{q}}(t) = \int d^3r \; \psi_{\mathbf{q}}^{*}(\mathbf{r},t) \hat{\Psi}(\mathbf{r}),
\end{equation}
and expanding the field operator on the complete set of time-dependent Bloch states 
\begin{equation}
\hat{\Psi}(\mathbf{r})= \sum\limits_{\mathbf{k}}  \psi_{\mathbf{k}}(\mathbf{r},t) \hat{a}_{\mathbf{k}}(t)  + \sum\limits_{l\neq1,\mathbf{k}}  \psi_{l,\mathbf{k}}(\mathbf{r},t) \hat{a}_{l,\mathbf{k}}(t) 
\end{equation} 
one obtains
\begin{equation}\label{eq:inblochstates}
\partial_t \hat{a}_{\mathbf{q}}(t) = \sum\limits_{\mathbf{k}} C_{\mathbf{k},\mathbf{q}}^{*}(t) \hat{a}_{\mathbf{k}}(t) + \sum\limits_{l \neq 1, \mathbf{k}} C^{*}_{l,\mathbf{k},\mathbf{q}}(t) \hat{a}_{l,\mathbf{k}}(t),
\end{equation}
where 
\begin{align}\label{eq:coefficients}
C_{\mathbf{k},\mathbf{q}}(t) = & \int d^3r\ \psi_{\mathbf{k}}^{*}(\mathbf{r},t) \frac{\partial }{\partial t} \psi_{\mathbf{q}}(\mathbf{r},t), \\
C_{l,\mathbf{k},\mathbf{q}}(t) = & \int d^3r\ \psi_{l,\mathbf{k}}^{*}(\mathbf{r},t) \frac{\partial }{\partial t} \psi_{\mathbf{q}}(\mathbf{r},t).
\end{align}
the first sum on the right-hand side of Eq.~(\ref{eq:inblochstates}) runs within the lowest energy band, while the second one runs over all the other bands. 
For ramping times that are long with respect to the inverse recoil frequency $\hbar/E_R$, one can neglect all inter-band couplings, i.e. $C_{l,\mathbf{k},\mathbf{q}}= 0$.
 Moreover, from the conservation of quasi-momentum we know that $C_{\mathbf{k},\mathbf{q}}(t) = \delta_{\mathbf{q} , \mathbf{k}}C_{\mathbf{q},\mathbf{q}}(t)$ (see Appendix~\ref{sec:symmetry}). The time derivative~\eqref{eq:single_lambda_derivative} of the number conserving operators is then
\begin{equation}
 \partial_t \left( \begin{array}{c}
\hat{\Lambda}_{\mathbf{q}}(t) \\
\hat{\Lambda}_{-\mathbf{q}}^{\dagger}(t)
\end{array}\right)
 = C_{\mathbf{q},\mathbf{q}}^{*}(t) 
\left( \begin{array}{c}
\hat{\Lambda}_{\mathbf{q}}(t) \\
\hat{\Lambda}_{-\mathbf{q}}^{\dagger}(t)
\end{array}\right),
\end{equation}
where we used the relations $C_{-\mathbf{q},-\mathbf{q}}(t) = C_{\mathbf{q},\mathbf{q}}^{*}(t)$ and $C_{\mathbf{0},\mathbf{0}}(t) = 0$~
\footnote{From the normalization of the Bloch states one can show that $\text{Re}\{ C_{\mathbf{q},\mathbf{q}}(t)\} = 0$. Moreover, $\psi_{l=1,\mathbf{0}}(\mathbf{r},t)$ is purely real.}
for the Bloch states in (\ref{eq:coefficients}). 

Gathering all the terms, the Heisenberg equation (\ref{eq:HeisenbergforB}) for $\hat{B}_{\bf q}(t)$ takes the form
\begin{align}
i\hbar \frac{d}{dt}\hat{B}_{\mathbf{q}}(t) & = \left[ \hbar \omega_{\mathbf{q}}(t) \sigma_z - i\hbar \Omega_{\mathbf{q}}(t) \sigma_x\right] \hat{B}_{\mathbf{q}}(t) \nonumber\\
& + i\hbar C_{\mathbf{q},\mathbf{q}}^{*}(t) \hat{B}_{\mathbf{q}}(t).
\label{eq:Bdot}
\end{align}
Notice that the second term in the above equation gives a global time-dependent phase factor, which can be removed by a gauge transformation. Finally, the solution of the Heisenberg equation (\ref{eq:HeisenbergforB}) can be cast in the simple form
\begin{equation}
\hat{B}_{\mathbf{q}}(t) = \left( \begin{array}{c}
\mathcal A_{\mathbf{q}}(t) \\ \mathcal B_{\mathbf{q}}(t)
\end{array}\right) \hat{b}_{\mathbf{q}}(0)
+ \left( \begin{array}{c}
\mathcal B_{\mathbf{q}}^{*}(t) \\ \mathcal A_{\mathbf{q}}^{*}(t)
\end{array}\right) \hat{b}^{\dagger}_{-\mathbf{q}}(0),
\end{equation}
where 
\begin{equation}\label{eq:one_AB_motion}
i\hbar \frac{d}{dt} 
\left( \begin{array}{c}
\mathcal A_{\mathbf{q}}(t) \\ \mathcal B_{\mathbf{q}}(t)
\end{array} 
\right) = 
\left( \begin{array}{cc}
\hbar \omega_{\mathbf{q}}(t) & -i\hbar \Omega_{\mathbf{q}}(t) \\
-i\hbar \Omega_{\mathbf{q}}(t) & -\hbar \omega_{\mathbf{q}}(t)
\end{array}
\right)
\left( \begin{array}{c}
\mathcal A_{\mathbf{q}}(t) \\ \mathcal B_{\mathbf{q}}(t)
\end{array} 
\right),
\end{equation}
with the initial conditions $\mathcal A_{\mathbf{q}}(0) = 1$ and $\mathcal B_{\mathbf{q}}(0) = 0$.

%-------------------------------------------------------------------------------------------
\subsection{The adiabaticity parameter}

During time evolution some Bogoliubov quasi-particles will be excited with respect to the quasi-particle vacuum state. The number quasi-particles created in mode $\mathbf{q}$ is determined by the coefficient $\mathcal B_{\mathbf{q}}(t)$:
\begin{equation}
n^{\rm ex}_{\mathbf{q}}(t) = \langle \Psi_{\rm Bog}(0) | \hat{b}^{\dagger}_{\mathbf{q}}(t) \hat{b}_{\mathbf{q}}(t) | \Psi_{\rm Bog}(0) \rangle = |\mathcal B_{\mathbf{q}}(t)|^2,
\end{equation}
where $| \Psi_{\rm Bog}(0) \rangle$ is the Bogoliubov vacuum state at time $t=0$. Excitations are avoided by suppressing the coupling between $\mathcal{A}_{\mathbf{q}}(t)$ and $\mathcal{B}_{\mathbf{q}}(t)$ in Eq.~\eqref{eq:one_AB_motion}, which brings us to the adiabaticity condition 
\begin{equation}\label{eq:adiabatic}
\alpha_{\mathbf{q}}(t) = \left|\frac{\hbar \Omega_{\mathbf{q}}(t)}{2 \hbar \omega_{\mathbf{q}}(t)}\right| \ll 1 ,\quad \forall {\mathbf{q}},
\end{equation}
where $\Omega_{\mathbf{q}}$ is the coupling (\ref{eq:one_omega}) and $\omega_{\mathbf{q}}$ is the Bogoliubov energy (\ref{eq:bogoliubov_energy}).
Since both $|\hbar \Omega_{\mathbf{q}}(t)|$ and $1/\hbar \omega_{\mathbf{q}}(t)$ are monotonically decreasing functions of $q$, the left-hand side of Eq.~\eqref{eq:adiabatic} reaches its maximum value for minimal quasi-momenta $|\mathbf{q}_{\rm min}|=q_{\rm min}$. 
We thus introduce the adiabaticity parameter $\alpha$ as:
\begin{eqnarray}\label{eq:single_adiabaticity_param}
 \alpha &=& \max\limits_{0 \leq t \leq t_{\rm ramp}}\alpha_{q_{\rm min}}(t), \\
 \alpha_{q_{\rm min}}(t) &\equiv&  \frac{\hbar}{4 \Delta E_{\mathbf{q}_{\rm min}}}\left|\frac{d}{dt}\left( \frac{\Delta E_{\mathbf{q}_{\rm min}}}{\hbar \omega_{\mathbf{q}_{\rm min}}}\right)\right|.
\end{eqnarray}

\subsubsection{Limit of large $N$}
By expanding Eq.~\eqref{eq:epsilonq}, \eqref{eq:bogoliubov_energy} and \eqref{eq:one_omega} for a small wave vector $q$ and non-zero interactions:
one obtains 
\begin{equation}
\alpha_{q_{\rm min}}(t) \stackrel{q_{\rm min}\to0}{=}  \frac{\hbar}{4J(t)}\left|\frac{d}{dt}\left( \frac{J(t)}{\hbar c(t)q_{\rm min}}Ê\right)\right|.
\end{equation}
The expression {of the} minimum wavenumber in the lattice
\begin{equation}
q_{\rm min} d = \frac{2\pi}{N^{1/3}},
\label{eq:qmin}
\end{equation}
gives a scaling $N^{1/3}$ to the adiabaticity parameter.
In the case of a linear ramp changing between $V_{\rm min}$ and $V_{\rm max}$ in a time $t_{\rm ramp}$ 
\begin{equation}
\label{eq:ramp}
V_0(t)=\frac{V_{\rm max}-V_{\rm min}}{t_{\rm ramp}}t+V_{\rm min},
\end{equation}
{taking $n=1$,} and using the approximated formulas (\ref{eq:GaussU})-(\ref{eq:GaussJ}) for a deep enough lattice,
we deduce the adiabatic time $t_{\rm adiab}$ such that the evolution is adiabatic for $t_{\rm ramp} \gg t_{\rm adiab}$,
\begin{align}
 t_{\rm adiab} = & \frac{\hbar}{E_R}  \; \frac{V_{\rm max} - V_{\rm min}}{E_R} \; \frac{N^{1/3}}{64(\sqrt{2}\pi ka_s)^{1/2}}
  \nonumber\\
 & \times \max\limits_{V_{\rm min} \leq V_0 \leq V_{\rm max}} \left(\frac{E_R}{V_0}\right)^{5/4} e^{\sqrt{\frac{V_0}{E_R}}}.
\end{align}
Note that the fact the formula gives a diverging time in the limit $V_0/E_R \to \infty$ (corresponding to $J\to0$) is not relevant here as 
our analysis is restricted to the superfluid regime were a condensate is present. 

\subsubsection{Ideal gas}
When interactions tends to zero, i.e. $U \to 0$ we cannot linearize the dispersion relation of quasi-particles. One rather has 
\begin{eqnarray}
\hbar \omega_{\mathbf{q}}(t) &\to& \Delta E_{\mathbf{q}}(t),\\
\Omega_{\mathbf{q}}(t)  &\to& 0 \,.
\end{eqnarray}
In the ideal gas regime all the particles occupy the instantaneous $\mathbf{q}=\mathbf{0}$ mode during time evolution and the evolution is always adiabatic. 
This effect is a consequence of quasi-momentum conservation, deriving from the fact the although the lattice height increases, the periodicity of the lattice is unchanged.

\subsection{Interpretation of the adiabatic parameter}
In this subsection we explicit the link between Eq.~\eqref{eq:adiabatic} and Eq.~\eqref{eq:adiabatic_cond2}, 
gaining some physical insight into our result.
Let us take the time derivative of the Bogoliubov Hamiltonian (\ref{eq:HBog}) 
\begin{align}
& i\hbar \frac{d}{dt} \hat{\mathcal{H}}_{\rm Bog}(t) = {i\hbar}{\frac{dH_0}{dt} + } {i\hbar}\sum\limits_{\mathbf{q}\neq \mathbf{0}}\frac{d(\hbar \omega_{\mathbf{q}}(t))}{dt} 
\left( \frac{1}{2} + \hat{b}^{\dagger}_{\mathbf{q}} \hat{b}_{\mathbf{q}} \right)\nonumber \\
&+ \hbar \omega_{\mathbf{q}}(t) \left[ i\hbar \left( \frac{d\hat{b}^{\dagger}_{\mathbf{q}}}{dt} \hat{b}_{\mathbf{q}} + 
 \hat{b}^{\dagger}_{\mathbf{q}} \frac{d\hat{b}_{\mathbf{q}}}{dt}\right)\right]. 
 \label{eq:derHBog}
\end{align}
The term in the first line of~\eqref{eq:derHBog} cannot change the number of quasi-particles and will play no role. For the term in square brackets in the second line, using Eq~\eqref{eq:Bdot} we find 
\begin{equation}
i\hbar \left( \frac{d\hat{b}^{\dagger}_{\mathbf{q}}}{dt} \hat{b}_{\mathbf{q}} + 
 \hat{b}^{\dagger}_{\mathbf{q}} \frac{d\hat{b}_{\mathbf{q}}}{dt}\right) = -i\hbar \Omega_{\mathbf{q}} \left( \hat{b}_{-\mathbf{q}} \hat{b}_{\mathbf{q}} + 
 \hat{b}^{\dagger}_{\mathbf{q}} \hat{b}^{\dagger}_{-\mathbf{q}}\right), 
\end{equation} 
showing that the time derivative of the Hamiltonian can couple the Bogoliubov ground state to states with two quasi-particles with opposite quasi-momenta
\begin{equation}
\langle 1:\mathbf{q},1:-\mathbf{q} |\hat{b}^{\dagger}_{\mathbf{q}} \hat{b}^{\dagger}_{-\mathbf{q}}|Ê0\rangle =1,
\end{equation}
the energy difference being $2\hbar \omega_{\mathbf{q}}$.
Using Eq.~\eqref{eq:adiabatic_cond2}, we then obtain the condition 
\begin{equation}
\alpha_{\mathbf{q}}(t)=\frac{\hbar |\Omega_{\mathbf{q}}| }{ 2\hbar \omega_{\mathbf{q}}} \ll 1,
\end{equation}
that coincides with the adiabaticity condition~\eqref{eq:adiabatic}.

%-------------------------------------------------------------------------------------------
\section{Adiabaticity criterion for two-components}
\label{sec:twocomp}

In this section we extend the reasoning presented in section \ref{sec:onecomp} to \revision{derive an adiabaticity condition for the two-component system,
including the case of different masses in the two components, corresponding to different isotopes \cite{WiemanRb8785} or different atomic species \cite{CataniK41Rb87,CornishCs133Yb172}.}

%-------------------------------------------------------------------------------------------
\subsection{Two-components Bose-Hubbard model}

We consider a two-component Bose-Einstein condensate in an optical lattice potential. \revision{As the atoms in the two components might have different masses, we introduce two tunneling 
parameters $J_a(t)$ and $J_b(t)$ of the Bose-Hubbard Hamiltonian, and two kinetic energies $\epsilon_{\mathbf{q},a}$, $\epsilon_{\mathbf{q},b}$ and $\Delta E_{\mathbf{q},a}$, $\Delta E_{\mathbf{q},b}$ defined as in equations 
(\ref{eq:J}), (\ref{eq:epsilonq}) and (\ref{eq:DeltaE}).
The interactions between atoms may also be different for different components.}
For convenience we introduce the interaction parameters 
\begin{equation}
U_{\sigma}(t) = \frac{a_{\sigma}}{a_s} U(t), \quad \mbox{with}\quad \sigma=a,b,ab,
\end{equation} 
with $a_{\sigma=a,b,ab}$ the $s$-wave scattering lengths respectively for two atoms
in the state $|a\rangle$, two atoms in the state $|b\rangle$, or one atom in $|a\rangle$ and the other in $|b\rangle$, where $U(t)$ is still defined by Eq.~\eqref{eq:u_definition}. 
\revision{We restrict here to the case of repulsive interactions $U_{\sigma}>0$ for $\sigma=a,b,ab$ and to the miscible regime $U_{ab}<\sqrt{U_aU_b}$ \cite{Fetter78,TLO,Nahm2005}.}
We note $N_a$ and $N_b$ the {average} number of atoms in each component, and $N=N_a+N_b$ the total number of atoms.

%-------------------------------------------------------------------------------------------
\subsection{Bogoliubov description for two-components}
The Bose-Hubbard Hamiltonian for the two-component system in the quasi-momentum representation takes the form
\begin{align}\label{eq:BoseHubbrdtwocomp}
& \hat{\mathcal{H}}_{2C}(t) = \sum\limits_{\mathbf{q}}  \left[ \revision{\epsilon_{\mathbf{q},a}(t)} \hat{c}^{\dagger}_{\mathbf{q},a}\hat{c}_{\mathbf{q},a} +  \revision{\epsilon_{\mathbf{q},b}(t)} \hat{c}^{\dagger}_{\mathbf{q},b}\hat{c}_{\mathbf{q},b}  \right] \nonumber\\
& + \sum\limits_{\mathbf{q}_1, \mathbf{q}_2, \mathbf{k}} \left[ \frac{U_a(t)}{2M}\hat{c}^{\dagger}_{\mathbf{q}_1 - \mathbf{k},a}\hat{c}^{\dagger}_{\mathbf{q}_2+\mathbf{k},a}\hat{c}_{\mathbf{q}_1,a}\hat{c}_{\mathbf{q}_2,a} \right. \nonumber\\
&\left.  \,\,\,\,\,\,\,\, \,\,\,\,\,\,\,\,\,\,\,\,\,\, 
+\frac{U_b(t)}{2M} \hat{c}^{\dagger}_{\mathbf{q}_1-\mathbf{k},b}\hat{c}^{\dagger}_{\mathbf{q}_2+\mathbf{k},b}\hat{c}_{\mathbf{q}_1,b}\hat{c}_{\mathbf{q}_2,b} \right.\nonumber \\
&\left . \,\,\,\,\,\,\,\, \,\,\,\,\,\,\,\,\,\,\,\,\,\,
+\frac{U_{ab}(t)}{M}\hat{c}^{\dagger}_{\mathbf{q}_1-\mathbf{k},a}\hat{c}^{\dagger}_{\mathbf{q}_2+\mathbf{k},b}\hat{c}_{\mathbf{q}_1,a}\hat{c}_{\mathbf{q}_2,b} \right],
\end{align}
{where $\hat{c}^{\dagger}_{\mathbf{q},\sigma}$ is the creation operator in the internal state $|\sigma \rangle$ and quasi-momentum $\mathbf{q}$.}
Similarly as in the single-component case, we introduce a set of number conserving operators 
\begin{eqnarray}
\hat{\Lambda}_{\mathbf{q},a} &=& \frac{1}{\sqrt{N_a}} \hat{c}_{\mathbf{0},a}^{\dagger}\hat{c}_{\mathbf{q},a} \, ,\\
\hat{\Lambda}_{\mathbf{q},b} &=& \frac{1}{\sqrt{N_b}} \hat{c}_{\mathbf{0},b}^{\dagger}\hat{c}_{\mathbf{q},b} \, .
\end{eqnarray}
In terms of the vector
\begin{equation}
\hat{\Lambda}_{2C,\mathbf{q}}= \left(  \hat{\Lambda}_{\mathbf{q},a},  \hat{\Lambda}_{-\mathbf{q},a}^{\dagger}, \hat{\Lambda}_{\mathbf{q},b},  \hat{\Lambda}_{-\mathbf{q},b}^{\dagger} \right)^T \,,
\end{equation}
the two-component Bogoliubov Hamiltonian (\ref{eq:BoseHubbrdtwocomp}), once quadratized, reads
\begin{equation}
\hat{\mathcal{H}}_{2C, \rm Bog}(t) = H_{2C,0}(t) + 
\frac{1}{2}\sum\limits_{\mathbf{q} \neq \mathbf{0}} 
\hat{\Lambda}_{2C,\mathbf{q}}^{\dagger}
\: \Sigma_z \mathcal{L}_{2C,\mathbf{q}} \;
\hat{\Lambda}_{2C,\mathbf{q}},
\label{eq:two_bogoliubov_matrix}
\end{equation}
where $\Sigma_z = \sigma_z \oplus \sigma_z$, and where the explicit form of the ground state energy $H_{2C,0}(t)$ and of the non-hermitian matrix $\mathcal{L}_{2C,\mathbf{q}}$, 
which are the two component equivalents of (\ref{eq:onecomp_H0}) and (\ref{eq:onecomp_L}) respectively,
is given in Appendix~\ref{ap:Ltwocomp}. 

Equivalently to \eqref{eq:bogoliubov_single}, the Bogoliubov Hamiltonian \eqref{eq:two_bogoliubov_matrix} can be diagonalized using {a} Bogoliubov transformation:
\begin{equation}\label{eq:bogol2C}
\hat{B}_{2C, \mathbf{q}}(t) = \mathcal{T}_{2C, \mathbf{q}}(t) \; \hat{\Lambda}_{2C, \mathbf{q}},
\end{equation}
{
where we have introduced the vector
\begin{equation}
\hat{B}_{2C, \mathbf{q}}(t) = (\hat{b}_{\mathbf{q},+}(t), \hat{b}^{\dagger}_{-\mathbf{q},+}(t), \hat{b}_{\mathbf{q},-}(t), \hat{b}^{\dagger}_{-\mathbf{q},-}(t))^T,
\label{eq:B2C}
\end{equation} 
$\hat{b}_{\mathbf{q},\pm}$ being annihilation operators of Bogoliubov quasi-particles, and the transformation matrix
}
\begin{equation}
\mathcal{T}_{2C, \mathbf{q}}(t)=
 \left(
 \begin{array}{cccc}
  \bar{u}_{\mathbf{q},+}^{a}(t) & -\bar{v}_{\mathbf{q},+}^{a}(t) & \bar{u}_{\mathbf{q},+}^{b}(t) & -\bar{v}_{\mathbf{q},+}^{b}(t) \\
  -\bar{v}_{\mathbf{q},+}^{a}(t) & \bar{u}_{\mathbf{q},+}^{a}(t) & -\bar{v}_{\mathbf{q},+}^{b}(t) & \bar{u}_{\mathbf{q},+}^{b}(t) \\
  \bar{u}_{\mathbf{q},-}^{a}(t) & -\bar{v}_{\mathbf{q},-}^{a}(t) & \bar{u}_{\mathbf{q},-}^{b}(t) & -\bar{v}_{\mathbf{q},-}^{b}(t) \\
  -\bar{v}_{\mathbf{q},-}^{a}(t) & \bar{u}_{\mathbf{q},-}^{a}(t) & -\bar{v}_{\mathbf{q},-}^{b}(t) & \bar{u}_{\mathbf{q},-}^{b}(t)
 \end{array}
\right)\,.
\end{equation}
As in the homogeneous case without a lattice \cite{leshouches,PhysRevA.67.023606,1751-8121-41-14-145005,SanchezPalencia}, there are two excitation branches labeled $\pm$:
\begin{align}
\hat{\mathcal{H}}_{\rm Bog}(t) & = H_{2C,0}(t) + \frac{1}{2}\sum\limits_{\mathbf{q} \neq \mathbf{0}}\sum\limits_{\sigma\in\{+,-\}} \hbar \omega_{\mathbf{q},\sigma}(t) \nonumber\\
& + \sum\limits_{\mathbf{q} \neq \mathbf{0}} \sum\limits_{\sigma\in\{+,-\}} \hbar \omega_{\mathbf{q},\sigma}(t) \hat{b}^{\dagger}_{\mathbf{q},\sigma}(t)\hat{b}_{\mathbf{q},\sigma}(t),
\end{align}
of energies
\begin{align}
&\hbar \omega_{\mathbf{q},\pm} =  \left[\frac{\hbar\omega_{\mathbf{q},a}^2 + \hbar\omega_{\mathbf{q},b}^2}{2} \right.\nonumber\\
&\left.\pm \sqrt{\frac{(\hbar\omega_{\mathbf{q},a}^2 - \hbar\omega_{\mathbf{q},b}^2)^2}{4} + 4\revision{\Delta E_{\mathbf{q},a}\Delta E_{\mathbf{q},b}} U_{ab}^2 n_a n_b} \,\right]^{1/2}
\label{eq:omegaq}
\end{align}
where $\hbar\omega_{\mathbf{q},a(b)}$ are defined as in the single component case~\eqref{eq:bogoliubov_energy}
\begin{eqnarray}
\hbar\omega_{\mathbf{q},a} &= & \sqrt{\revision{\Delta E_{\mathbf{q},a}} (\revision{\Delta E_{\mathbf{q},a}} + 2U_a n_a)}, \\
\hbar\omega_{\mathbf{q},b} &= & \sqrt{\revision{\Delta E_{\mathbf{q},b}}  (\revision{\Delta E_{\mathbf{q},b}} + 2U_b n_b)},
\end{eqnarray}
with $n_\sigma = N_\sigma/M$. {For single-particle unit filling $M=N$, $n_a$ and $n_b$ represent the fraction of atoms in component $a$ and $b$ respectively.}
The corresponding Bogoliubov amplitudes which appear in the transformation~\eqref{eq:bogol2C} are \cite{1751-8121-41-14-145005}
\begin{align}\label{eq:BogolAmplitudes2C}
\left(
 \begin{array}{c}
  \bar{u}_{\mathbf{q},\pm}^{a} \\[3mm]
  \bar{v}_{\mathbf{q},\pm}^{a} \\[3mm]
  \bar{u}_{\mathbf{q},\pm}^{b} \\[3mm]
  \bar{v}_{\mathbf{q},\pm}^{b} 
 \end{array}
\right)
= &
\left(
 \begin{array}{c}
  2U_{ab}\sqrt{n_a n_b}\revision{\Delta E_{\mathbf{q},b}} (\revision{\Delta E_{\mathbf{q},a}}+ \hbar \omega_{\mathbf{q},\pm}) \\[3mm]
  2U_{ab}\sqrt{n_a n_b} \revision{\Delta E_{\mathbf{q},b}}  (\revision{\Delta E_{\mathbf{q},a}} - \hbar \omega_{\mathbf{q},\pm}) \\[3mm]
  (\hbar \omega_{\mathbf{q},\pm}^2 - \hbar \omega_{\mathbf{q},a}^2)  (\revision{\Delta E_{\mathbf{q},b}}+ \hbar \omega_{\mathbf{q},\pm})\\[3mm]
  (\hbar \omega_{\mathbf{q},\pm}^2 - \hbar \omega_{\mathbf{q},a}^2)  (\revision{\Delta E_{\mathbf{q},b}} - \hbar \omega_{\mathbf{q},\pm})
 \end{array}
\right)
\chi_{\mathbf{q},\pm},
\end{align}
with the normalization coefficient $\chi_{\mathbf{q},\pm}$ given by
\begin{align}
&\chi_{\mathbf{q},\pm} =  \left[4 \revision{\Delta E_{\mathbf{q},b}}  \hbar \omega_{\mathbf{q},\pm} \right.\nonumber\\
&\left.\times \left(4 \revision{\Delta E_{\mathbf{q},a}\Delta E_{\mathbf{q},b}} U_{ab}^2 n_a n_b + (\hbar \omega_{\mathbf{q},\pm}^2 - \hbar \omega_{\mathbf{q},a}^2)^2 \right) \right]^{-1/2}
\end{align}
\revision{is chosen to have $\sum_{\sigma=a,b} |\bar{u}_{\mathbf{q},\pm}^{\sigma}|^2-|\bar{v}_{\mathbf{q},\pm}^{\sigma}|^2=1$.}

%-------------------------------------------------------------------------------------------
\subsection{Heisenberg equation of motion}
Following the procedure and arguments presented for the single component case in section \ref{sec:onecomp},
the Heisenberg equation of motion for the operator $\hat{B}_{2C,\mathbf{q}}(t)$, defined in Eq.~\eqref{eq:B2C} takes the form
\begin{align}
& i\hbar \frac{d}{dt} \hat B_{2C,\mathbf{q}}(t) = \nonumber\\
&\,\,\,\,\,\left[ \hbar \omega_{\mathbf{q},+}(t) (\sigma_z \oplus 0_{2 \times 2}) 
 + \hbar \omega_{\mathbf{q},-}(t) (0_{2\times  2} \oplus \sigma_z) \right] \hat B_{2C,\mathbf{q}}(t) \nonumber\\
& + i\hbar \frac{\mathcal{T}_{2C,\mathbf{q}}(t)}{dt} \mathcal{T}_{2C,\mathbf{q}}^{-1}(t) \hat B_{2C,\mathbf{q}}(t)
+ i\hbar C_{\mathbf{q},\mathbf{q}}^{*} \hat B_{2C,\mathbf{q}}(t).
\label{eq:two_heisenberg}
\end{align}
The last term on the right-hand side of Eq.~\eqref{eq:two_heisenberg} gives a global time-dependent phase factor, which can be removed by a suitable gauge transformation.
In general, the coupling term in the Eq.~\eqref{eq:two_heisenberg} is
\begin{align}
& \frac{\mathcal{T}_{2C,\mathbf{q}}(t)}{dt} \mathcal{T}_{2C,\mathbf{q}}^{-1}(t) = \nonumber\\[3mm]
& \left(
 \begin{array}{cccc}
  0 & -\Omega_{\mathbf{q},+}(t) & -\Omega_{\mathbf{q},1}(t) & -\Omega_{\mathbf{q},2}(t) \\
  -\Omega_{\mathbf{q},+}(t) & 0 & -\Omega_{\mathbf{q},2}(t) & -\Omega_{\mathbf{q},1}(t) \\
 \revision{-} \Omega_{\mathbf{q},1}(t) & -\Omega_{\mathbf{q},2}(t) & 0 & -\Omega_{\mathbf{q},-}(t) \\
  -\Omega_{\mathbf{q},2}(t) & \revision{-} \Omega_{\mathbf{q},1}(t) & -\Omega_{\mathbf{q},-}(t) & 0
 \end{array}
\right),
\end{align}
with the following couplings 
\begin{align}
\Omega_{\mathbf{q},\pm}(t) & = \sum\limits_{\sigma=a,b} \bar{u}_{\mathbf{q},\pm}^{\sigma}(t) \frac{d}{dt} \bar{v}_{\mathbf{q},\pm}^{\sigma}(t) - \bar{v}_{\mathbf{q},\pm}^{\sigma}(t) \frac{d}{dt} \bar{u}_{\mathbf{q},\pm}^{\sigma}(t) \label{eq:coup0}\\
\Omega_{\mathbf{q},1}(t) & = \sum\limits_{\sigma=a,b} \bar{v}_{\mathbf{q},-}^{\sigma}(t) \frac{d}{dt} \bar{v}_{\mathbf{q},+}^{\sigma}(t) - \bar{u}_{\mathbf{q},-}^{\sigma}(t) \frac{d}{dt} \bar{u}_{\mathbf{q},+}^{\sigma}(t), 
 \label{eq:coup1}\\
\Omega_{\mathbf{q},2}(t) & = \sum\limits_{\sigma=a,b} \bar{u}_{\mathbf{q},-}^{\sigma}(t) \frac{d}{dt} \bar{v}_{\mathbf{q},+}^{\sigma}(t) - \bar{v}_{\mathbf{q},-}^{\sigma}(t) \frac{d}{dt} \bar{u}_{\mathbf{q},+}^{\sigma}(t),
 \label{eq:coup2}
\end{align}
\revision{
that are explicitly calculated in appendix~\ref{app:sd_and_Coupl}, in equations (\ref{eq:coupl0G}-\ref{eq:coupl2G}). 
In the general case, the amplitudes $\hat{b}_{\mathbf{q},\pm}(t)$ and $\hat{b}_{-\mathbf{q},\pm}(t)$ are coupled in the time dependent optical lattice.
In particular one has:
\begin{equation}
i\hbar \frac{d}{dt}
\left(
 \begin{array}{c}
  \hat{b}_{\mathbf{q},+}(t) \\
  \hat{b}_{\mathbf{q},-}(t) \\
  \hat{b}^\dagger_{-\mathbf{q},+}(t) \\
  \hat{b}^\dagger_{-\mathbf{q},-}(t) 
 \end{array}
\right)
 = {\cal M}(t)
\left(
 \begin{array}{c}
  \hat{b}_{\mathbf{q},+}(t) \\
  \hat{b}_{\mathbf{q},-}(t) \\
  \hat{b}^\dagger_{-\mathbf{q},+}(t) \\
  \hat{b}^\dagger_{-\mathbf{q},-}(t) 
 \end{array}
\right) 
\label{eq:expansion}
\end{equation}
with the evolution matrix ${\cal M}(t)$
\begin{equation}
 {\cal M}(t)=\left(
 \begin{array}{cccc}
  \hbar \omega_{\mathbf{q},+} & -i\hbar \Omega_{\mathbf{q},1} & -i\hbar \Omega_{\mathbf{q},+} & -i\hbar \Omega_{\mathbf{q},2}\\
  -i\hbar \Omega_{\mathbf{q},1} & \hbar \omega_{\mathbf{q},-} & -i\hbar \Omega_{\mathbf{q},2} & -i\hbar \Omega_{\mathbf{q},-}\\
-i\hbar \Omega_{\mathbf{q},+} & -i\hbar \Omega_{\mathbf{q},2}& -\hbar \omega_{\mathbf{q},+} & -i\hbar \Omega_{\mathbf{q},1}  \\
  -i\hbar \Omega_{\mathbf{q},2} & -i\hbar \Omega_{\mathbf{q},-} & -i\hbar \Omega_{\mathbf{q},1} & -\hbar \omega_{\mathbf{q},-} 
 \end{array}
\right) 
\label{eq:matrixM}
\end{equation}
By using the symplectic symmetry of the matrix ${\cal M}$, that is of the form 
$
 {\cal M}(t)=\left(
 \begin{array}{cc}
  A & B\\
  -B^\ast & -A^\ast
 \end{array}
\right) 
$
we can write
\begin{widetext}
\begin{equation}
\left(
 \begin{array}{c}
  \hat{b}_{\mathbf{q},+}(t) \\
  \hat{b}_{\mathbf{q},-}(t) \\
  \hat{b}^\dagger_{-\mathbf{q},+}(t) \\
  \hat{b}^\dagger_{-\mathbf{q},-}(t) 
 \end{array}
\right) = 
\left(
 \begin{array}{c}
  \mathcal{A}_{\mathbf{q}, +}(t) \\
  \mathcal{C}_{\mathbf{q}, +}(t) \\
  \mathcal{B}_{\mathbf{q}, +}(t) \\
  \mathcal{D}_{\mathbf{q}, +}(t) 
 \end{array}
\right) \hat{b}_{\mathbf{q},+}(0) 
+
\left(
 \begin{array}{c}
  \mathcal{C}_{\mathbf{q}, -}(t) \\
  \mathcal{A}_{\mathbf{q}, -}(t) \\
  \mathcal{D}_{\mathbf{q}, -}(t) \\
  \mathcal{B}_{\mathbf{q}, -}(t) 
 \end{array}
\right) \hat{b}_{\mathbf{q},-}(0) 
+
\left(
 \begin{array}{c}
  \mathcal{B}^\ast_{\mathbf{q}, +}(t) \\
  \mathcal{D}^\ast_{\mathbf{q}, +}(t) \\
  \mathcal{A}^\ast_{\mathbf{q}, +}(t) \\
  \mathcal{C}^\ast_{\mathbf{q}, +}(t) 
 \end{array}
\right) \hat{b}^\dagger_{-\mathbf{q},+}(0) 
+
\left(
 \begin{array}{c}
  \mathcal{D}^\ast_{\mathbf{q}, -}(t) \\
  \mathcal{B}^\ast_{\mathbf{q}, -}(t) \\
  \mathcal{C}^\ast_{\mathbf{q}, -}(t) \\
  \mathcal{A}^\ast_{\mathbf{q}, -}(t) 
 \end{array}
\right) \hat{b}^\dagger_{-\mathbf{q},-}(0)
\label{eq:cal}
\end{equation}
\end{widetext}
where the evolution equation for each column vector in the right hand side of equation (\ref{eq:expansion}) is governed by the matrix ${\cal M}(t)$
\begin{equation}
i\hbar \frac{d}{dt}
\left(
 \begin{array}{c}
  \mathcal{A}_{\mathbf{q}, +}(t)\\
  \mathcal{C}_{\mathbf{q}, +}(t) \\
  \mathcal{B}_{\mathbf{q}, +}(t) \\
  \mathcal{D}_{\mathbf{q}, +}(t) 
 \end{array}
\right)
 = {\cal M}(t)
\left(
 \begin{array}{c}
  \mathcal{A}_{\mathbf{q}, +}(t) \\
  \mathcal{C}_{\mathbf{q}, +}(t) \\
  \mathcal{B}_{\mathbf{q}, +}(t) \\
  \mathcal{D}_{\mathbf{q}, +}(t) 
 \end{array}
\right) 
\label{eq:evol_amplitudes}
\end{equation}
etc., with and initial conditions $\mathcal{A}_{\mathbf{q},\pm}(0) = 1$ and $\mathcal{B}_{\mathbf{q},\pm}(0)=\mathcal{C}_{\mathbf{q},\pm}(0) =\mathcal{D}_{\mathbf{q},\pm}(0)=0$.

%-------------------------------------------------------------------------------------------
\subsection{Two components adiabaticity parameter}

\subsubsection{General case}
According to the expansion (\ref{eq:expansion}), and assuming that no excitations were initially present, the number of Bogoliubov quasi-particle in mode $\mathbf{q}$ created by the ramp is
\begin{eqnarray}
n_{\mathbf{q}}^{\rm ex}(t) &=& \langle \hat{b}^\dagger_{\mathbf{q},+}(t) \hat{b}_{\mathbf{q},+}(t) \rangle + \langle \hat{b}^\dagger_{\mathbf{q},-}(t) \hat{b}_{\mathbf{q},-}(t) \rangle \\
	&=& \sum_{\sigma=\pm }|\mathcal{B}_{\mathbf{q},\sigma}(t) |^2 + |\mathcal{D}_{\mathbf{q},\sigma}(t) |^2 .
\end{eqnarray}
In order to minimize the total amount of excitations during the time evolution, one needs to reduce the off-diagonal couplings in the evolution equation of the amplitudes (\ref{eq:evol_amplitudes}). 
This leads to the adiabaticity conditions~:
\begin{eqnarray}
\left|\frac{\hbar\Omega_{\mathbf{q},+}(t)}{2 \hbar\omega_{\mathbf{q},+}(t)}\right| \ll 1 \quad ; \quad \left|\frac{\hbar\Omega_{\mathbf{q},-}(t)}{2 \hbar\omega_{\mathbf{q},-}(t)}\right| &\ll &1 \label{eq:adiaba_pm} \\
\left| \frac{\Omega_{\mathbf{q},1}(t)}{\hbar \omega_{\mathbf{q},+}(t) - \hbar \omega_{\mathbf{q},-}(t)} \right| &\ll& 1 \\
\left| \frac{\Omega_{\mathbf{q},2}(t)}{\hbar \omega_{\mathbf{q},+}(t) + \hbar \omega_{\mathbf{q},-}(t)} \right| &\ll& 1 \,.
\end{eqnarray}
}

\subsubsection{Equal masses}
\revision{Let us now concentrate on the case of equal masses, corresponding to two internal states of the same atomic species.
In this case, with $\Delta E_{\mathbf{q},a}=\Delta E_{\mathbf{q},b}\equiv\Delta E_{\mathbf{q}}$, $J_a(t)=J_b(t)\equiv J(t)$, 
the couplings $\Omega_{\mathbf{q},1,2}(t)$ vanish as $U_a(t)$, $U_b(t)$ and $U_{ab}(t)$ share the same time dependence. 
It follows that the Bogoliubov modes of the ``+" and ``-" branch 
evolve independently and we are left with the adiabaticity conditions in equation (\ref{eq:adiaba_pm}), with}
\begin{equation}\label{eq:two_coupling}
\Omega_{\mathbf{q},\pm }(t) = \frac{1}{2}\frac{d}{dt}\log \left( \frac{\Delta E_{\mathbf{q}}(t)}{\hbar \omega_{\mathbf{q},\pm}(t)}\right)\,.
\end{equation}
The adiabatic  {conditions are} most stringent at the minimal quasi-momenta $|\mathbf{q}_{\rm min}|=q_{\rm min}$~\eqref{eq:qmin}, and one introduces the two adiabaticity parameters $\alpha_{\pm}$ 
\begin{eqnarray}
\label{eq:two_adiabaticity_cond}
 \alpha_{\pm} &=& \max\limits_{0 \le t \le t_{\rm ramp}} \alpha_{q_{\rm min,\pm}}(t), \\
 \alpha_{q_{\rm min},\pm}(t) &\equiv&  \frac{\hbar}{4 \Delta E_{\mathbf{q}_{\rm min}}} \left|\frac{d}{dt}\left( \frac{\Delta E_{\mathbf{q}_{\rm min}}}{\hbar \omega_{\mathbf{q}_{\rm min},\pm}}\right)\right|,
\end{eqnarray}
corresponding to the two excitation branches in the two-component system.
In order for the evolution to be adiabatic the conditions $\alpha_{\pm} \ll 1$ must be satisfied.

In the limit of a large particle number $N\to \infty$, $q_{\rm min}\to0$ and single-particle unit filling $n_a + n_b=1$
\begin{equation}
 \alpha_{q_{\rm min},\pm} \stackrel{q_{\rm min}\to0}{=}  \frac{\hbar}{4J(t)} \left|\frac{d}{dt} \left( \frac{J(t)}{\hbar c_{\pm}(t) q_{\rm min}} \right) \right|,
\end{equation}
where $c_{\pm}(t)$ are the sound velocities of the two excitation branches defined as follows 
\begin{align}
c_{\pm}^2(t) & = \frac{1}{2}\left[c_a^2(t) + c_b^2(t) \right.\nonumber\\
& \left. \pm \sqrt{(c_b^2(t) - c_a^2(t))^2 + 4 \frac{U_{ab}^2(t)}{U_a(t) U_b(t)}c_a^2(t)c_b^2(t) } \right],
\label{eq:bogol_two_sound}\\
c_a(t) & = \frac{{d}}{\hbar}\sqrt{2J(t) U_a(t) n_a}, \\
c_b(t) & = \frac{{d}}{\hbar}\sqrt{2J(t) U_b(t) n_b}.
\end{align}

\section{Adiabatic time}
\label{sec:equalmasses}

In this section, \revision{for the case of equal masses corresponding to two hyperfine states of the same atomic species,}
we derive the expression of the adiabatic time for a linear ramp and we investigate the influence of a density imbalance between two components.

\begin{figure*}[]
\includegraphics[width=0.322\linewidth,clip=true]{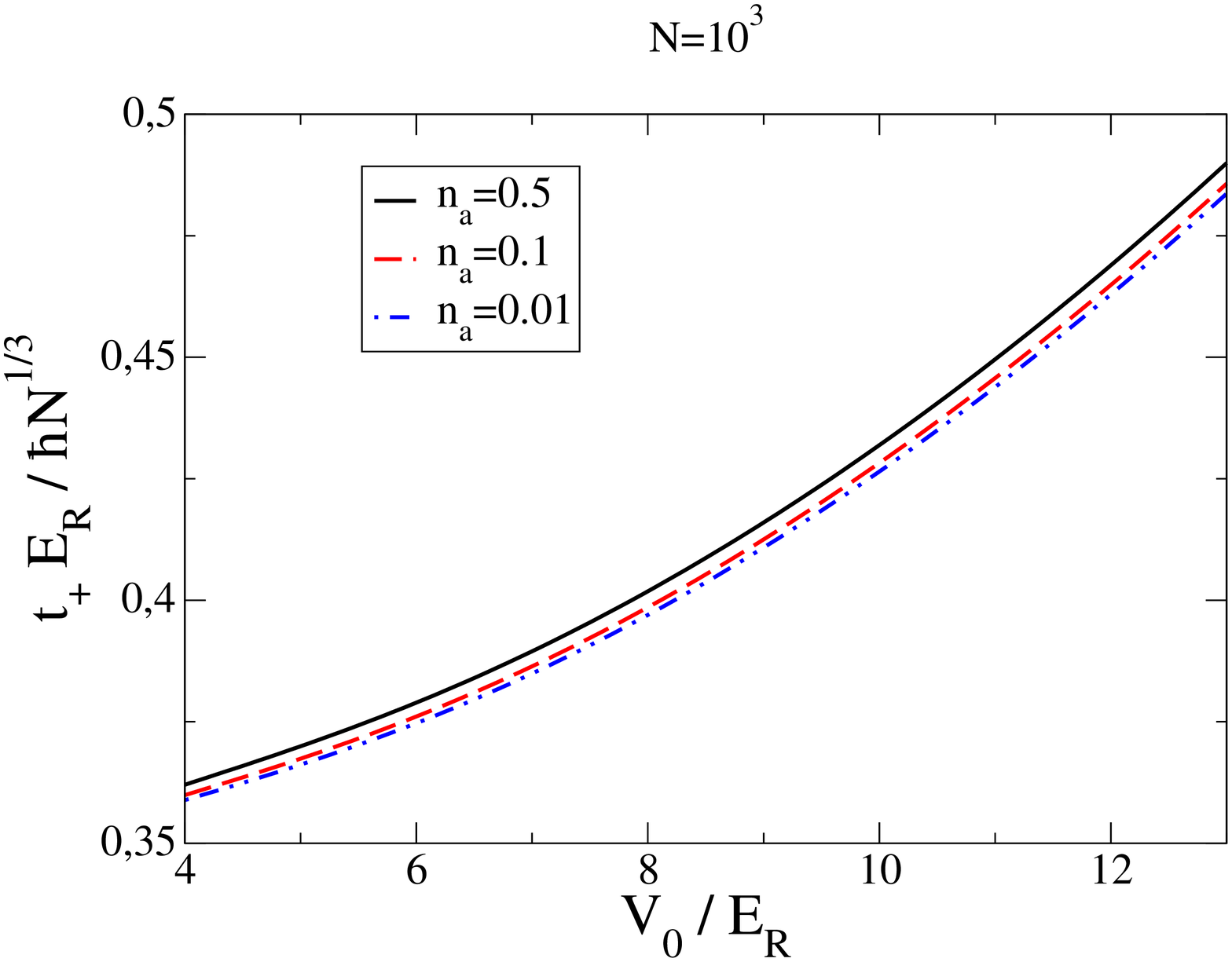}\:\:\:\:\includegraphics[width=0.322\linewidth,clip=true]{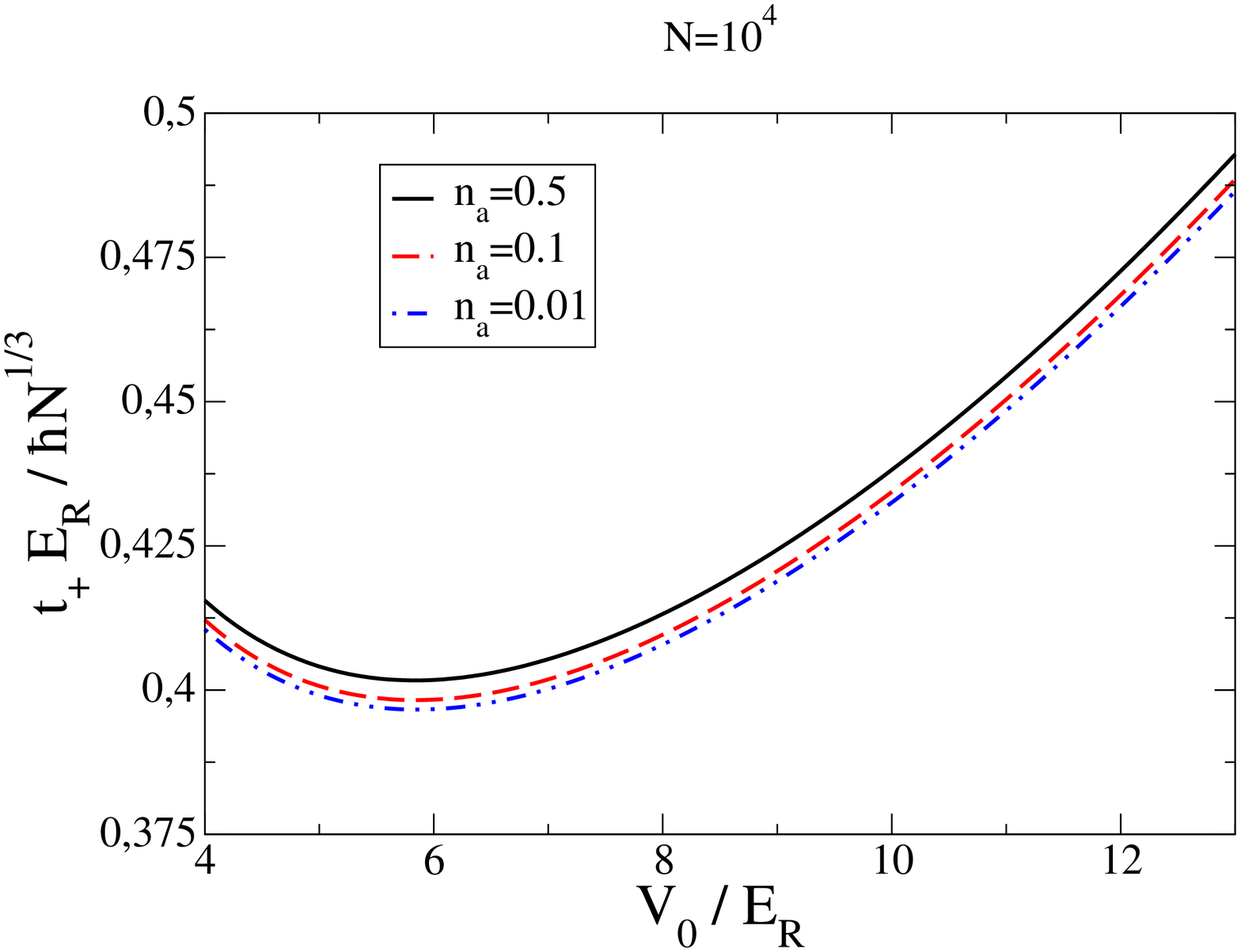}\:\:\:\:\includegraphics[width=0.322\linewidth,clip=true]{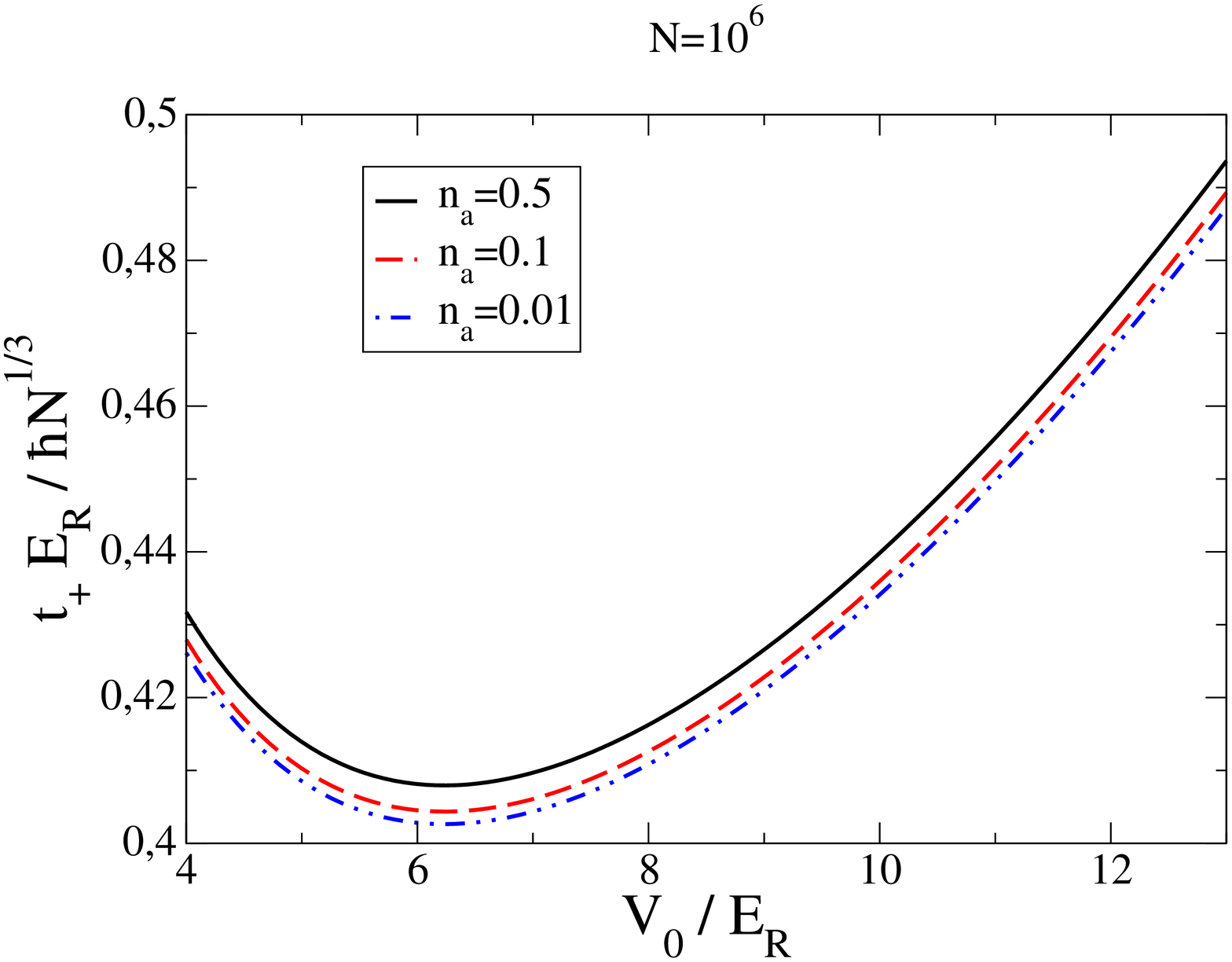}
\includegraphics[width=0.32\linewidth,clip=true]{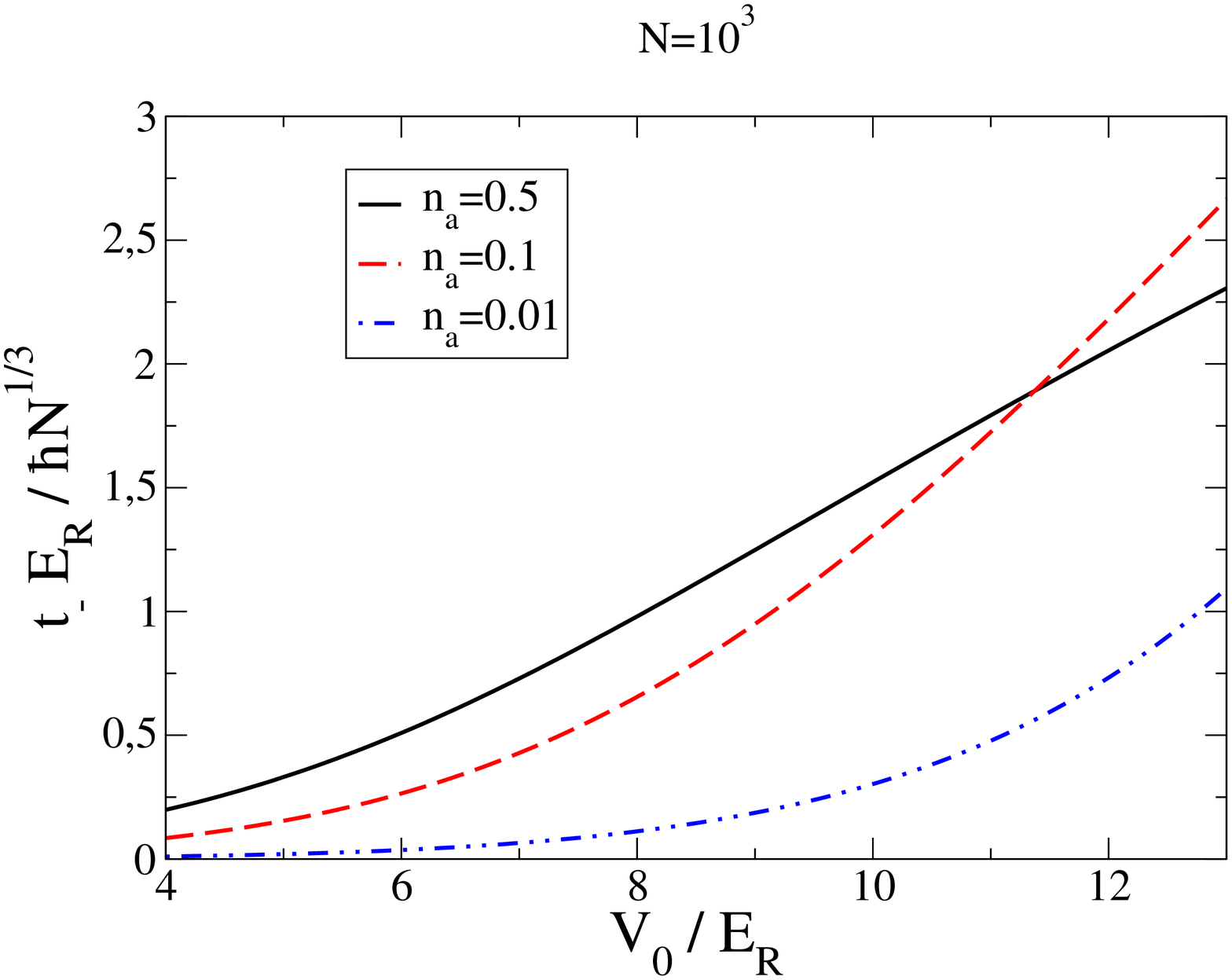}\:\:\:\:\includegraphics[width=0.32\linewidth,clip=true]{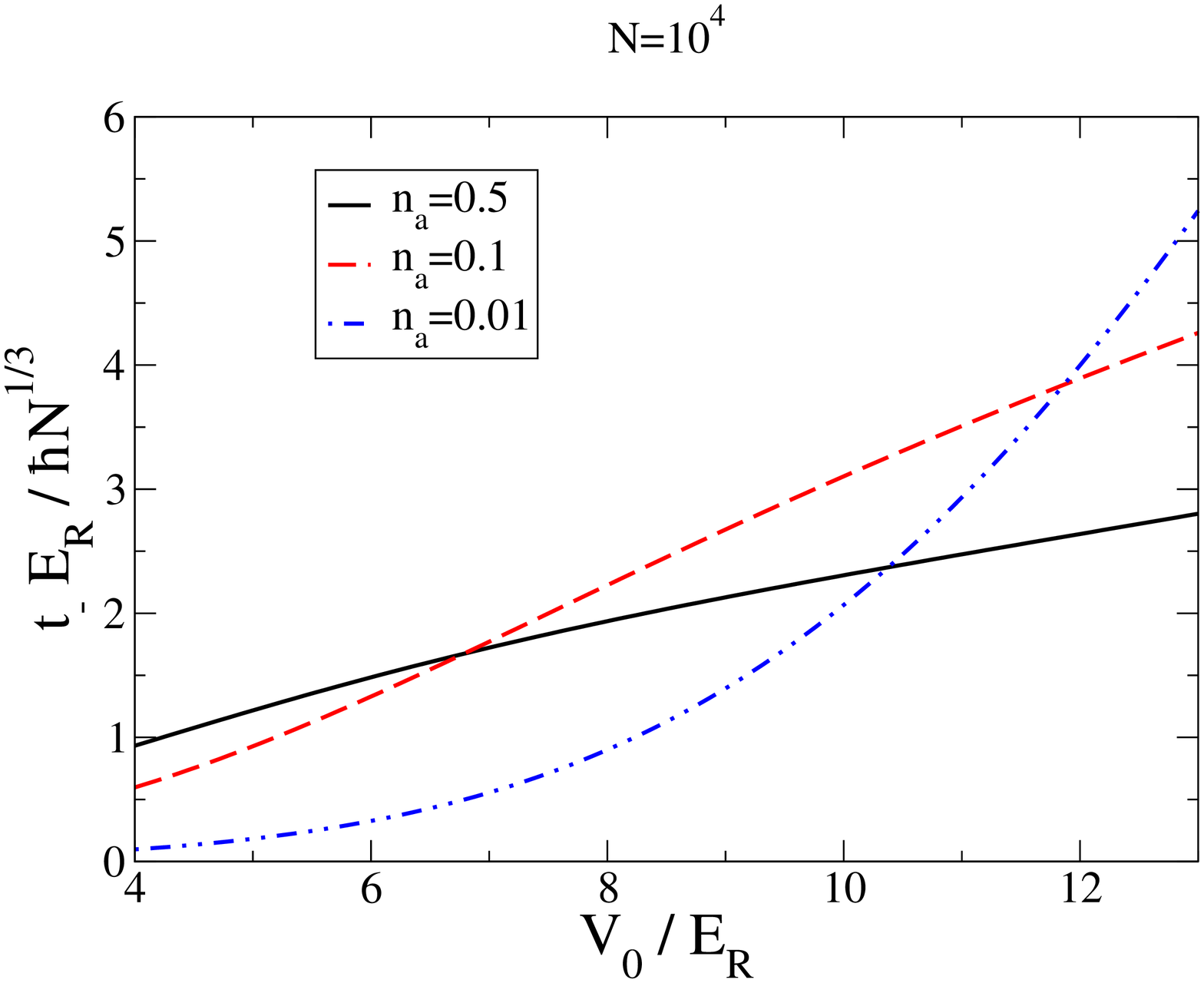}\:\:\:\:\includegraphics[width=0.32\linewidth,clip=true]{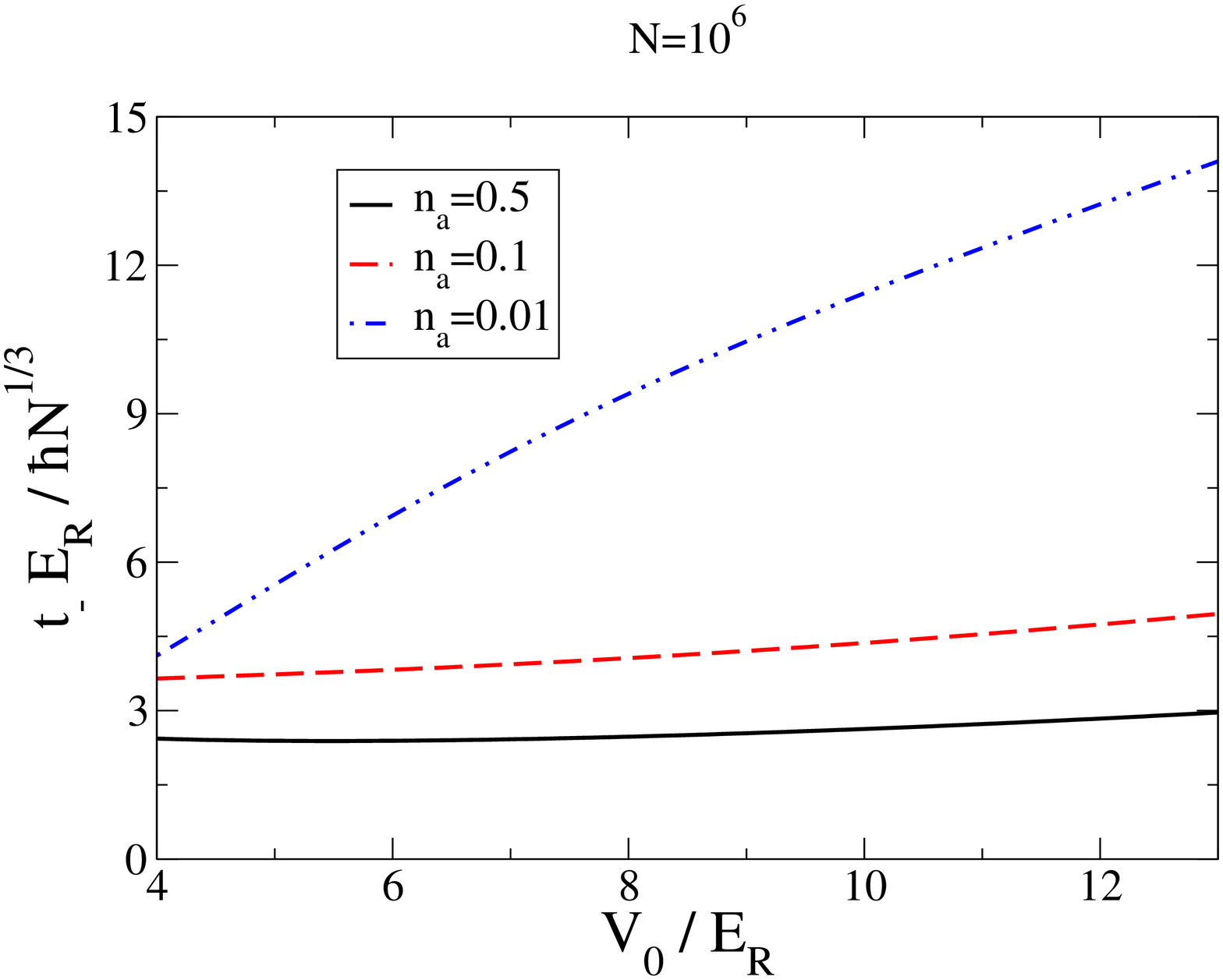}
\caption{(Color online) Times $t_{+}(V_0)$ (top row) and $t_{-}(V_0)$ (bottom row) in equation (\ref{eq:tpmV0}) as a function of $V_0/E_R$, for three atom numbers: $N=10^3$, $N=10^4$ and $N=10^6$ from the left to the right. In each panel we show the result for different fractions of atoms in component $a$: $n_a=0.5$ (black line), $n_a=0.1$ (red dashed line) and $n_a=0.01$ (blue dash-dotted line). In each case the maximum of $t_{\pm}$ is reached for the largest value of $V_0/E_R$. The plots where generated {expanding the kinetic energy term $\Delta E_{\mathbf{q}_{\rm min}}$ for a low quasi-momentum $dq_{\rm min}\ll1$ and} using the approximate formulas (\ref{eq:GaussU})-(\ref{eq:GaussJ}) for $J$ and $U_{a,b,ab}$ with scattering lengths $a_{aa} = a_{bb} = 100.4 r_0$ and $a_{ab} = 95.0 r_0$ where $r_0$ is the Bohr radius,
\revision{corresponding to the hyperfine sates $F=\pm1$, $m_F=\mp1$ of ${}^{87}$Rb atoms with a Feshbach-tuned interspecies scattering length $a_{ab}$ \cite{Oberthaler2010,AtomicCrystal}.} The lattice wavelengths is $\lambda=800$nm.}
\label{fig:figV0}
\end{figure*}

For a linear ramp~\eqref{eq:ramp}, 
by changing variables from $t$ to $V_0$ in~\eqref{eq:two_adiabaticity_cond}, the adiabaticity condition is $t_{\rm ramp}\gg t_{\rm adiab, \pm} $, where
 \begin{align}
 &t_{\rm adiab,\pm} = \max\limits_{ V_{\rm min} \leq V_0 \leq V_{\rm max}} t_{\pm}(V_0) \label{eq:two_adiab}, \\
 &t_{\pm}(V_0)  =\frac{\hbar (V_{\rm max}-V_{\rm min})}{4 \Delta E_{\mathbf{q}_{\rm min}}(V_0)} \left| \frac{d}{dV_0}\left( \frac{\Delta E_{\mathbf{q}_{\rm min}}(V_0)}{\hbar \omega_{\mathbf{q}_{\rm min},\pm}(V_0)}\right) \right|.
 \label{eq:tpmV0}
 \end{align}
Let us consider a situation in which $V_{\rm min}= 4 E_R$, $V_{\rm max}= 13 E_R$ and the fraction $n_a$ of atoms in component $a$ is varied from 0 to 1, with $n_b=1-n_a$.

In Fig.~\ref{fig:figV0} we plot the times $t_{\pm}(V_0)$ in Eq.~\eqref{eq:tpmV0} as a function of $V_0/E_R$, for different atom numbers and fractions $n_a$.
Within the selected range of $V_0$, the maximum of $t_{\pm}$ is reached for $V_0=V_{\rm max}$.
The maximum of the ``minus" branch (bottom row) is always larger than that of the ``plus" branch (top row), hence setting the minimal time scale for adiabatic evolution.

\begin{figure*}[]
\includegraphics[width=0.43\linewidth,clip=true]{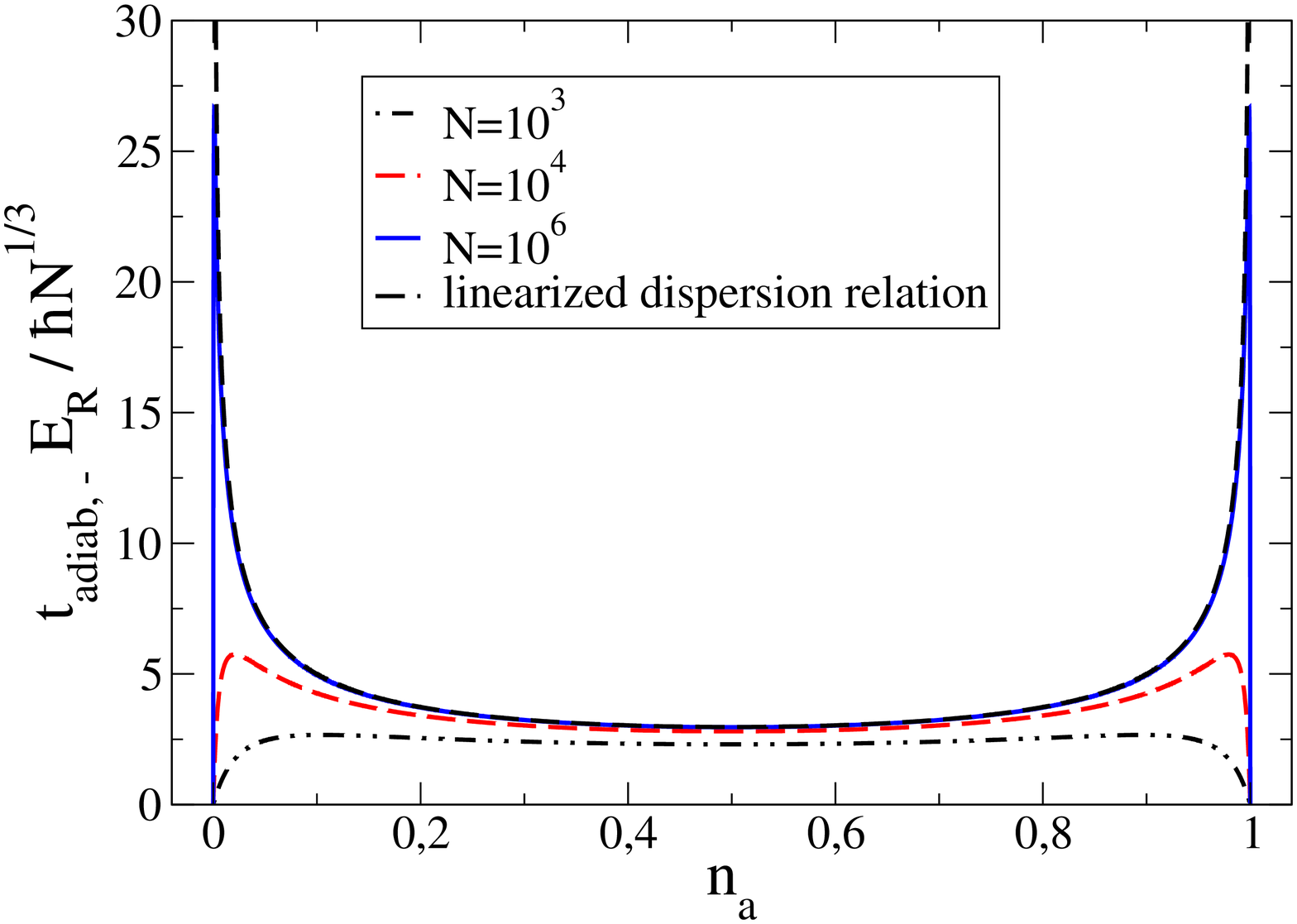}\: \: \includegraphics[width=0.43\linewidth,clip=true]{TOUSNb2_log.eps}
\caption{(Color online) Adiabatic time $t_{\rm adiab, -}$ as a function of the fraction $n_a$ of atoms in component $a$, for different atom numbers, in linear scale (left panel) or in semi-log scale (right panel).
The solid lines are from Eq.~\eqref{eq:two_adiab}: $N=10^3$ (black dash-dotted line), $N=10^4$ (red dashed line), $N=10^6$ (blue solid line), while the black dashed line is from the approximate expression \eqref{eq:two_approx} that holds in the large $N$ limit. 
The plots where generated {expanding the kinetic energy term $\Delta E_{\mathbf{q}_{\rm min}}$ for a low quasi-momentum $dq_{\rm min}\ll1$ and} using the approximate formulas (\ref{eq:GaussU})-(\ref{eq:GaussJ}) for $J$ and $U_{a,b,ab}$ with scattering lengths $a_{aa} = a_{bb} = 100.4 r_0$ and $a_{ab} = 95.0 r_0$
where $r_0$ is the Bohr radius. The lattice wavelengths is $\lambda=800$nm.}
\label{fig:tadiab}
\end{figure*}

In Fig.~\ref{fig:tadiab} we plot the adiabatic time $t_{\rm adiab,-}$ given by Eq.~\eqref{eq:two_adiab} as a function of $n_a$ for different atom numbers (colored solid lines).
The black dashed line is the approximation for large $N$:
 \begin{align}
 & t_{\rm adiab,\pm} \stackrel{q_{\rm min}\to0}{=}  \max\limits_{ V_{\rm min} \leq V_0 \leq V_{\rm max}}  t_{\pm}^{\rm lin}(V_0), \label{eq:two_approx} \\
 & t_{\pm}^{\rm lin}(V_0) = \frac{\hbar (V_{\rm max}-V_{\rm min})}{4 J(V_0)} \left| \frac{d}{dV_0} \left( \frac{J(V_0)}{\hbar c_{\pm}(V_0) q_{\rm min}}\right) \right|,
 \label{eq:tpmV0_lin}
 \end{align}

\subsection{Discussion of the results}
To gain physical insight in the low energy excitations  {involved in the adiabaticty condition for the two component system}, let us first consider 
the homogeneous case without the lattice~\cite{leshouches,SanchezPalencia}. A clear physical picture is obtained by linearizing the coupled Gross-Pitaevskii equations 
\begin{eqnarray}
i\hbar \partial_t \psi_a &=& \left[ -\frac{\hbar^2\Delta}{2m} + U_{a} |\psi_a|^2 + U_{ab} |\psi_b|^2 \right] \psi_a, \label{eq:linGPEa}\\
i\hbar \partial_t \psi_b &=& \left[ -\frac{\hbar^2\Delta}{2m} + U_{b} |\psi_b|^2 + U_{ab} |\psi_a|^2 \right] \psi_b, \label{eq:linGPEb}
\end{eqnarray}
for the fields of the two components
\begin{equation}
\psi_a(\mathbf{r},t)=\sqrt{N_a}\phi_a(\mathbf{r},t) \quad \psi_b(\mathbf{r},t)=\sqrt{N_b}\phi_b(\mathbf{r},t),
\end{equation}
around the uniform solutions $\psi_a=\sqrt{n_a}$, $\psi_b=\sqrt{n_b}$ with chemical potentials $\mu_a=n_a U_{a}+n_b U_{ab}$,
$\mu_b=n_b U_{b}+n_a U_{ab}$, where $\phi_a=\phi_b=1/\sqrt{V}$ are the stationary condensate wave functions, $n_a=N_a/V$, $n_b=N_b/V$ are the uniform atomic densities,
and $U_{\sigma}=\frac{4\pi \hbar^2 a_{\sigma}}{m}$ for $\sigma=a,b,ab$ are the interaction constants. Linearization of Eq.~(\ref{eq:linGPEa})-(\ref{eq:linGPEb}) gives
\begin{eqnarray}
i\hbar \partial_t \delta \psi_a = &-&\frac{\hbar^2\Delta}{2m} \delta \psi_a + n_a U_{a} (\delta \psi_a + \delta \psi_a^\ast) \nonumber \\
&+&\sqrt{n_a n_b} U_{ab} (\delta \psi_b + \delta \psi_b^\ast), \label{eq:lina} \\
i\hbar \partial_t \delta \psi_b = &-&\frac{\hbar^2\Delta}{2m} \delta \psi_b + n_b U_{b} (\delta \psi_b + \delta \psi_b^\ast) \nonumber \\
&+&\sqrt{n_a n_b} U_{ab} (\delta \psi_a + \delta \psi_a^\ast), \label{eq:linb}
\end{eqnarray}
One then looks for eigenmodes in the form of plane waves
\begin{equation}
\binom{\delta \psi_{{\sigma}}}{\delta \psi_{{\sigma}}^\ast} = \sum_{\mathbf{q}\neq \mathbf{0}} 
\binom{\bar{u}^{{\sigma}}_\mathbf{q}}{\bar{v}^{{\sigma}}_\mathbf{q}} e^{i(\mathbf{q}\cdot\mathbf{r}-\omega_\mathbf{q} t)}.
\end{equation}
{where $\sigma=a,b$.}
Note that in the density-phase point of view where 
$\psi_a=\sqrt{\rho_a} e^{i\theta_a}$ and $\psi_b=\sqrt{\rho_b} e^{i\theta_b}$ we obtain
\begin{eqnarray}
\delta \rho^{{\sigma}}_\mathbf{q} &=& \frac{1}{2} (\bar{u}^{{\sigma}}_\mathbf{q}+\bar{v}^{{\sigma}}_\mathbf{q}),  \\
\delta \theta^{{\sigma}}_\mathbf{q} &=&  \frac{1}{2i}  \frac{\bar{u}^{{\sigma}}_\mathbf{q}-\bar{v}^{{\sigma}}_\mathbf{q}}{\sqrt{n_{{\sigma}}}}\,.
\end{eqnarray}
The linearized equations {(\ref{eq:lina})-(\ref{eq:linb})} decouple when taking the sum and difference. One then finds that 
\begin{equation}
\bar{u}^a_\mathbf{q}-\bar{v}^a_\mathbf{q} = \frac{\hbar \omega_\mathbf{q}}{\Delta E_\mathbf{q}} (\bar{u}^a_\mathbf{q}+\bar{v}^a_\mathbf{q}), 
\end{equation}
and similarly for $b$, where $\Delta E_\mathbf{q} = \hbar^2 \mathbf{q}^2/2m$, is the kinetic energy, and the sum combination satisfies the eigenvalue problem
\begin{equation}
\left\{ \Delta E_\mathbf{q} \left[ 1- \left( \hbar \frac{\omega_\mathbf{q}}{\Delta E_\mathbf{q}}\right)^2 \right]\mathds{I} +2 \mathcal{M}  \right\} 
\binom{\bar{u}^a_\mathbf{q}+\bar{v}^a_\mathbf{q}}{\bar{u}^b_\mathbf{q}+\bar{v}^b_\mathbf{q}} =0,
\end{equation}
with {$\mathds{I}$ the identity matrix}  and 
\begin{equation}
\mathcal{M}=\binom{\:\:\: n_a U_{a} \quad \quad \sqrt{n_a n_b} \, U_{ab}}{\sqrt{n_a n_b} \, U_{ab} \quad \quad n_b U_{b}\:\:\:}.
\end{equation}
The eigenvalues of $\mathcal{M}$ are
\begin{equation}
\lambda_{\pm}=\frac{1}{2}\left[ U_{a}n_a+U_bn_b \pm \sqrt{(U_{a}n_a-U_bn_b)^2+4n_an_bU_{ab}^2}\right],
\end{equation}
and the two eigenenergies take the form 
\begin{equation}
\hbar \omega_{\mathbf{q},\pm} = \sqrt{\Delta E_\mathbf{q} \left[ \Delta E_\mathbf{q} +  2 \lambda_{\pm} \right]}.
\label{eq:omegaq_hom}
\end{equation}
Looking for the eigenvectors of the 2x2 matrix $\mathcal{M}$, one finds 
the ratio between the density perturbations in the two components for the two excitation branches 
\begin{align}
\frac{\delta \rho_{\pm}^b}{\delta \rho_{\pm}^a}&=\frac{\bar{u}^b_{\mathbf{q},\pm}+\bar{v}^b_{\mathbf{q},\pm}}{\bar{u}^a_{\mathbf{q},\pm}+\bar{v}^a_{\mathbf{q},\pm}} = \frac{\lambda_{\pm}-U_an_a}{U_{ab}\sqrt{n_an_b}} \nonumber \\
&=  \mbox{sgn}(\lambda_{\pm}-U_an_a)   \left(\frac{\lambda_{\pm}-U_an_a}{\lambda_{\pm}-U_bn_b} \right)^{1/2}.\label{eq:bella}
\end{align}
In the limit of a small wave vector $\mathbf{q}$, they correspond to sound waves with the speed of sound given by
\begin{equation}
c_\pm^{\rm homo}=\sqrt{\frac{\lambda_\pm}{m}}.
\label{eq:bogol_two_sound_homo}
\end{equation}

\medskip
{The relations in this subsection, derived for a homogeneous system,} hold {\it in the lattice}, provided one reinterprets the kinetic energy term $\Delta E_\mathbf{q}$ according to Eq.~\eqref{eq:DeltaE}. In particular Eq.~\eqref{eq:omegaq_hom} for the spectrum coincides with Eq.~\eqref{eq:omegaq}, Eq.~\eqref{eq:bella} can be easily deduced by the modal functions~\eqref{eq:BogolAmplitudes2C}, and Eq.~\eqref{eq:bogol_two_sound} for the speed of sound reduces to~\eqref{eq:bogol_two_sound_homo} by performing the substitution $\frac{1}{m} \to \frac{2J(t)d^2}{\hbar^2}$ that is the low energy equivalent of the correspondence 
(\ref{eq:DeltaE}) for kinetic energy. Provided {$U_b n_b\neq0$}, we can rewrite in the lattice  
\begin{align}
&c_\pm^2 =\frac{d^2J(t)U_b(t)n_b}{\hbar^2} \times F \label{eq:c2simple}, \\
&F= \left[ \left(1+\frac{U_a n_a}{U_b n_b}\right) \pm \sqrt{\left(1-\frac{U_a n_a}{U_b n_b}\right)^2+4\frac{n_a}{n_b}\frac{U_{ab}^2}{U_b^2}}\; \right].
\label{eq:qua}
\end{align}
Note that when $U_a(t)$ and $U_b(t)$ have the same {dependence on $V_0$}, the factor $F$ in~\eqref{eq:c2simple} gets out of the derivative in~\eqref{eq:tpmV0_lin}, and gives the dependence
of the adiabatic time on the ratios between the atom numbers and scattering lengths in the two components.
For the linear ramp {one indeed has}
\begin{equation}
t_\pm^{\rm lin}=\frac{\hbar N^{1/3}(V_{\rm max}-V_{\rm min})}{8\pi \sqrt{n_b F}}\frac{1}{J}\frac{d}{dV_0}\sqrt{\frac{J}{{U_b}}},
\label{eq:gulu}
\end{equation}
which also shows that, {for a linearized dispersion relation,} $t_{\rm adiab,-}$ is always larger than $t_{\rm adiab,+}$.

\subsubsection{\it Symmetric case} 
For $U_a=U_b$ and $n_a=n_b{=1/2}$, the factor $\sqrt{n_b F}$ in the denominator of the {$t_\pm^{\rm lin}$} (\ref{eq:gulu}) is equal to ${(1\pm U_{ab}/U_a)^{1/2}}$ showing the 
divergence of $t_-^{\rm lin}$ when $U_{ab} \to U_a$. 
Equation (\ref{eq:bella}) shows that the ``minus" solution corresponds to a situation in which the two component oscillate out-of-phase while the ``plus" solution 
corresponds to in-phase oscillations
\begin{eqnarray}
\delta \rho_{-}^a &=&-\delta \rho_{-}^b, \\
\delta \rho_{+}^a &=&\delta \rho_{+}^b. 
\end{eqnarray}
The out-of phase solution, whose speed of sound $c_-$
tends to zero as $U_{ab}$ approaches $U_a$ from below, announces the demixing instability for $U_{ab}^2>U_aU_b$.

\subsubsection{\it Asymmetric case} 
In the asymmetric case, {with $U_a=U_b$ but $n_a \ll n_b$}, to the lowest order in $n_a$, that is zero order, one has
\begin{eqnarray}
\hbar \omega_{\mathbf{q},-} &=& \Delta E_\mathbf{q}, \label{eq:asymb}\\
\hbar \omega_{\mathbf{q},+} &=& \sqrt{\Delta E_\mathbf{q}(\Delta E_\mathbf{q} + 2 U_b n_b) }. \label{eq:asyma}
\end{eqnarray}
The ``minus" branch $\hbar \omega_{\mathbf{q},-}$ does not describe a sound wave, but
is purely quadratic for small ${\bf q}$, see (\ref{eq:epsilonq})~\footnote{In the Hartree-Fock (HF) limit, where
\begin{equation}
\hbar \omega_{\mathbf{q}} \stackrel{HF}{=}  \Delta E_\mathbf{q} + \Delta_{\rm mf} -\mu,
\end{equation}
with $\Delta_{\rm mf}$ the mean field energy shift and $\mu$ the chemical potential,
the result (\ref{eq:asymb}) can be interpreted as the compensation for the minority component $a$ between the mean field shift due to interactions with $b$: 
$\Delta_{\rm mf}^a=U_{ab}n_b$ and the chemical potential $\mu_a=U_{ab}n_b$, giving
\begin{equation}
\hbar \omega_{\mathbf{q},-} \stackrel{HF}{=}  \Delta E_\mathbf{q},
\end{equation}
while in the majority component $b$ one has $\Delta_{\rm mf}^b=2U_{b}n_b$ (the factor 2 summing the Hartree and the Fock contribution) 
and $\mu_b=U_{b}n_b$ giving as for a single component
\begin{equation}
\hbar \omega_{\mathbf{q},+} \stackrel{HF}{=}  \Delta E_\mathbf{q} + U_{b}n_b\,.
\end{equation}
}. Correspondingly, for $n_a=0$ (and symmetrically for $n_a=1$) $\Omega_{\mathbf{q},-}=0$ in equation (\ref{eq:two_coupling}) and the adiabatic time is zero as 
shown in {the right panel of} figure \ref{fig:tadiab}. We note however that the adiabatic time significantly increases where $0<n_a\ll n_b$. This feature can be simply explained in the large $N$ limit
where, using equations {(\ref{eq:qua})} and (\ref{eq:gulu}), one gets
\begin{equation}
\frac{(t_-^{\rm lin})_{n_a\ll n_b}}{(t_-^{\rm lin})_{n_a=n_b}} \stackrel{n_a\to0}{\sim} \frac{1}{\sqrt{ 2\left(1+\frac{U_{ab}}{U_b} \right)}}\; \frac{1}{\sqrt{n_a}} \,.
\end{equation} 

\section{Conclusions}
\label{sec:concl}
We found the adiabaticity condition when an optical lattice is raised in a uniform two-components Bose-Einstein condensate at single particle unit filling. 
We concentrated our analysis in the superfluid regime {and we used} the time dependent Bogoliubov approach.
We find that the excitations that can brake adiabaticity are pairs of Bogoliubov excitations (sound waves) with opposite quasi-momenta, where, in the symmetric case  
the two components oscillate out of phase. The scaling of the adiabatic time with the system size is $N^{1/3}$, given by the minimal quasi momentum that can be excited in the lattice. 
For large atom numbers we show that the adiabatic time is significantly larger in the strongly asymmetric case {$n_a\ll n_b$} with respect to the symmetric case $n_a=n_b$, the ratio scaling as 
{$1/\sqrt{n_a}$}.

\section{Acknowledgments}
This work was supported by the Polish National Science Center Grants
DEC-2015/18/E/ST2/00760 and by the CNRS PICS-7403. 
D.K. acknowledges the financial support of the French Government (BGF) and the French Embassy in Poland. 

\appendix
\section{Implications of discrete translational symmetry \label{sec:symmetry}}
The single particle Hamiltonian \eqref{eq:single_particle_hamiltonian} is invariant under discrete translation $\mathbf{R}$ along any lattice vector. 
It follows that translation operator $\hat{T}_{\mathbf{R}}$ commutes with the Hamiltonian i.e. $[\hat{h}(t), \hat{T}_{\mathbf{R}}] = 0$, at any time. Complete set of commuting observables theorem implies that both operators share the same set of eigenstates. The eigenstates of the operator $\hat{T}_{\mathbf{R}}$ are labeled with discrete quasi-momentum $\mathbf{q}$ values \footnote{The discrete nature of quasi-momentum is a consequence of Born-von K\'{a}rm\'{a}n boundary conditions in the finite size system.}
\begin{equation}
\label{eq:translation_operator}
\hat{T}_{\mathbf{R}} \phi_{\mathbf{q}}(\mathbf{r},t) = e^{i \mathbf{q} \cdot \mathbf{R}} \phi_{\mathbf{q}}(\mathbf{r},t),
\end{equation}
with eigenvalue $e^{i \mathbf{q} \cdot \mathbf{R}}$. Equation~\eqref{eq:translation_operator} suggests the following form of the eigenstate 
\begin{equation}
\phi_{\mathbf{q}}(\mathbf{r},t) = e^{i\mathbf{q} \cdot \mathbf{r}} u_{\mathbf{q}}(\mathbf{r},t),
\end{equation}
where function $u_{\mathbf{q}}(\mathbf{r},t)$ has the same periodicity as the lattice. Thus, single-particle Hamiltonian eigenstates are of the form $\psi_{n,\mathbf{q}}(\mathbf{r},t) = e^{i\mathbf{q} \cdot \mathbf{r}} u_{n,\mathbf{q}}(\mathbf{r},t)$, with additional band index $n$. This is a statement of the Bloch theorem.

In a process where only the height of the potential varies in time and the periodicity of the lattice stays constant the time-evolution operator $\hat{U}(t)$ defined as
\begin{equation}
\hat{U}(t) = \mathcal{T} e^{-i\int\limits_{0}^{t}ds\ \hat{h}(s)},
\end{equation}
where $\mathcal{T}$ denotes time-ordering operator, also commutes with $\hat{T}_{\mathbf{R}}$. Consequently, the time-evolving state $\psi(\mathbf{r},t) = \hat{U}(t) \psi_{n,\mathbf{q}}(\mathbf{r})$ will remain an eigenstate of the translation operator with the same eigenvalue as the Bloch state $\psi_{n,\mathbf{q}}(\mathbf{r})$. In other words, quasi-momentum is conserved during time evolution. It follows from
\begin{equation}
\hat{T}_{\mathbf{R}} \hat{U}(t) \psi_{n,\mathbf{q}}(\mathbf{r}) = \hat{U}(t) \hat{T}_{\mathbf{R}} \psi_{n,\mathbf{q}}(\mathbf{r}) = e^{i\mathbf{q} \cdot \mathbf{R}} \hat{U}(t) \psi_{n,\mathbf{q}}(\mathbf{r}).
\end{equation}
Although the quasi-momentum needs to be conserved, it does not exclude situation where other bands may be populated. In general one can write
\begin{equation}
\psi(\mathbf{r},t) = \sum\limits_{m=1} c_{m}(t) \psi_{m,\mathbf{q}}(\mathbf{r},t),
\end{equation}
with $c_{m}(0) = \delta_{m,n}$.

Another consequence of translational invariance is the vanishing {of the} coupling term
\begin{equation}
C_{n,\mathbf{k}, m, \mathbf{q}}(t) = \int d^3 r\ \psi^{*}_{n,\mathbf{k}}(\mathbf{r},t) \frac{\partial }{\partial t} \psi_{m,\mathbf{q}}(\mathbf{r},t)
\end{equation}
when $\mathbf{k} \neq \mathbf{q}$. This can be shown once we consider the identity 
\begin{equation}
C_{n,\mathbf{k},m,\mathbf{q}}(t) = \frac{\int d^3 r\ \psi^{*}_{n,\mathbf{k}}(\mathbf{r},t) \frac{\partial \hat{h}(t)}{\partial t} \psi_{m,\mathbf{q}}(\mathbf{r},t)}{E_{m,\mathbf{q}}(t) - E_{n,\mathbf{k}}(t)}.
\end{equation}
and quasi-momentum conservation
\begin{align}
\hat{T}_{\mathbf{R}} \frac{\partial \hat{h}(t)}{\partial t} \psi_{m,\mathbf{q}}(\mathbf{r},t) & = \frac{\partial \hat{h}(t)}{\partial t}  \hat{T}_{\mathbf{R}} \psi_{m,\mathbf{q}}(\mathbf{r},t) \nonumber\\
& = e^{i\mathbf{q}\cdot \mathbf{R}} \frac{\partial \hat{h}(t)}{\partial t} \psi_{m,\mathbf{q}}(\mathbf{r},t). \label{eq:ham_time}
\end{align}
Eq.~\eqref{eq:ham_time} shows that
\begin{equation}
\frac{\partial \hat{h}(t)}{\partial t} \psi_{m,\mathbf{q}}(\mathbf{r},t) = \sum\limits_{l=1} c_{l}(t) \psi_{l,\mathbf{q}}(\mathbf{r},t).
\end{equation}
This result combined with orthogonality of Bloch states gives {a} non-zero coupling term only if {the} quasi-momentum of both states is the same.

\section{Explicit formulas for the two-component system case}\label{ap:Ltwocomp}

The non-hermitian matrix $\mathcal{L}$ is defined as follows
\begin{widetext}
\begin{equation}
\mathcal{L}_{2C,\mathbf{q}} = 
\left( 
 \begin{array}{cccc}
 \revision{\Delta E_{\mathbf{q},a}(t)} + U_a(t) n_a & U_a(t) n_a & U_{ab}(t)\sqrt{n_a n_b} & U_{ab}(t)\sqrt{n_a n_b} \\[4mm]
 -U_a(t) n_a & -[\revision{\Delta E_{\mathbf{q},a}(t)} + U_a(t) n_a ] & -U_{ab}(t)\sqrt{n_a n_b} & -U_{ab}(t)\sqrt{n_a n_b} \\[4mm]
 U_{ab}(t)\sqrt{n_a n_b} & U_{ab}(t)\sqrt{n_a n_b} & \revision{\Delta E_{\mathbf{q},b}(t)} + U_b(t) n_b  & U_b(t) n_b \\[4mm]
 -U_{ab}(t)\sqrt{n_a n_b} & -U_{ab}(t)\sqrt{n_a n_b} & -U_b(t) n_b & -[\revision{\Delta E_{\mathbf{q},b}(t)} + U_b(t) n_b]
 \end{array}
\right).
\end{equation}
The energy introduced in Eq.~\eqref{eq:two_bogoliubov_matrix} takes the form
\begin{align}
 H_{2C,0}(t) = & N_a \left[\epsilon_{\mathbf{0},\revision{a}}(t) + \frac{U_a(t)n_a}{2}\right] + N_b \left[\epsilon_{\mathbf{0},\revision{b}}(t) + \frac{U_b(t)n_b}{2} \right] + \frac{U_{ab}(t)}{M}N_a N_b 
 - \frac{1}{2}\sum\limits_{\mathbf{q} \neq \mathbf{0}} \revision{ \sum\limits_{\sigma=a,b}} \left[  \revision{\Delta E_{\mathbf{q},\sigma}(t)  + U_\sigma(t)n_\sigma }\right].
\end{align}
\end{widetext}

\section{\revision{Couplings among Bogoliubov modes}}
\label{app:sd_and_Coupl}
\revision{
It is convenient to introduce the sum and differences of the Bogoliubov modal functions ~\eqref{eq:BogolAmplitudes2C} 
\begin{equation}
s_{\mathbf{q},\pm}^\sigma=u_{\mathbf{q},\pm}^\sigma+v_{\mathbf{q},\pm}^\sigma \quad ; \quad d_{\mathbf{q},\pm}^\sigma=u_{\mathbf{q},\pm}^\sigma-v_{\mathbf{q},\pm}^\sigma 
\end{equation}
with $\sigma=a,b$, and express them using the parameters $x_{\mathbf{q},\pm}$ 
\begin{equation}
x_{\mathbf{q},\pm}=\frac{\hbar\omega_{\mathbf{q},\pm}^2 - \hbar\omega_{\mathbf{q},a}^2}{\sqrt{4\Delta E_{\mathbf{q},a}\Delta E_{\mathbf{q},b} U_{ab}^2 n_a n_b} } \,; \quad x_{\mathbf{q},+} \, x_{\mathbf{q},-}=-1
\label{eq:xpm}
\end{equation}
One has
\begin{eqnarray}
s_{\mathbf{q},\pm}^a&=&\sqrt{\frac{\Delta E_{\mathbf{q},a}}{\hbar \omega_\pm}} \sqrt{\frac{1}{1+x_{\mathbf{q},\pm}^2}}  \label{eq:sa} \\ 
d_{\mathbf{q},\pm}^a&=&\sqrt{\frac{\hbar \omega_\pm}{\Delta E_{\mathbf{q},a}}} \sqrt{\frac{1}{1+x_{\mathbf{q},\pm}^2}}  \label{eq:da} \\ 
s_{\mathbf{q},\pm}^b&=& x_{\mathbf{q},\pm} \sqrt{\frac{\Delta E_{\mathbf{q},b}}{\hbar \omega_\pm}} \sqrt{\frac{1}{1+x_{\mathbf{q},\pm}^2}}  \label{eq:sb} \\ 
d_{\mathbf{q},\pm}^b&=& x_{\mathbf{q},\pm} \sqrt{\frac{\hbar \omega_\pm}{\Delta E_{\mathbf{q},b}}} \sqrt{\frac{1}{1+x_{\mathbf{q},\pm}^2}}  \label{eq:db} \,.
\end{eqnarray}
Using equations (\ref{eq:sa})-(\ref{eq:db}), and equation (\ref{eq:coup0})-(\ref{eq:coup2}) we obtain
\begin{eqnarray}
2 \Omega_{\mathbf{q},\pm } &=& \frac{1}{1+x^2_{\mathbf{q},\pm}}\frac{d}{dt}\log \left( \frac{\Delta E_{\mathbf{q},a}}{\hbar \omega_{\mathbf{q},\pm}}\right) \nonumber \\
&+& \frac{x_{\mathbf{q},\pm}^2}{1+x^2_{\mathbf{q},\pm}}\frac{d}{dt}\log \left( \frac{\Delta E_{\mathbf{q},b}}{\hbar \omega_{\mathbf{q},\pm}}\right) \label{eq:coupl0G}\\
-2( \Omega_{\mathbf{q},1}+\Omega_{\mathbf{q},2} ) &=& \sqrt{\frac{\hbar \omega_{\mathbf{q},+}}{\hbar \omega_{\mathbf{q},-}}} \sqrt{\frac{1}{(1+x^2_{\mathbf{q},-})(1+x^2_{\mathbf{q},+})}}  \nonumber \\
&\times& \frac{d}{dt}\log \left( \frac{\Delta E_{\mathbf{q},b}}{\Delta E_{\mathbf{q},a}}  \frac{|x_{\mathbf{q},-}|}{|x_{\mathbf{q},+}|}  \right) \label{eq:coupl1G}\\
-2( \Omega_{\mathbf{q},1}-\Omega_{\mathbf{q},2} ) &=& \sqrt{\frac{\hbar \omega_{\mathbf{q},-}}{\hbar \omega_{\mathbf{q},+}}} \sqrt{\frac{1}{(1+x^2_{\mathbf{q},-})(1+x^2_{\mathbf{q},+})}}  \nonumber \\
&\times& \frac{d}{dt}\log \left( \frac{\Delta E_{\mathbf{q},a}}{\Delta E_{\mathbf{q},b}}  \frac{|x_{\mathbf{q},-}|}{|x_{\mathbf{q},+}|}  \right) \,.\label{eq:coupl2G}
\end{eqnarray}

}
\bibliography{revised2}

\end{document}